\documentclass[12pt]{iopart}
\usepackage{iopams}
\usepackage{graphicx}
\expandafter\let\csname equation*\endcsname\relax
\expandafter\let\csname endequation*\endcsname\relax
\usepackage{amsmath}
\usepackage{amssymb}
\usepackage{cite}

\usepackage{xcolor}

\begin{document}

\title[First hitting times of random walks on random regular graphs]
{Analytical results for the distribution of first hitting times 
of random walks on random regular graphs
}

\author{Ido Tishby, Ofer Biham and Eytan Katzav}
\address{Racah Institute of Physics, 
The Hebrew University, Jerusalem 9190401, Israel.}
\eads{\mailto{ido.tishby@mail.huji.ac.il}, \mailto{biham@phys.huji.ac.il}, 
\mailto{eytan.katzav@mail.huji.ac.il}}

\begin{abstract}

We present analytical results for the distribution of
first hitting times of
random walks (RWs) on random regular graphs (RRGs) of 
degree $c \ge 3$ and a finite size $N$.
Starting from a random initial node at time $t=1$, 
at each time step $t \ge 2$ an RW hops randomly 
into one of the $c$ neighbors of its previous node.
In some of the time steps the RW may hop into a yet-unvisited node
while in other time steps it may 
revisit a node that has already been visited before.
The first time at which the RW enters a node that has already been visited before is
called the first hitting time
or the first intersection length.
The first hitting event may take place either by backtracking (BT) to the previous node or by
retracing (RET), namely stepping into a node which has been visited two or more
time steps earlier.
We calculate the tail distribution 
$P( T_{\rm FH} > t )$
of first hitting (FH) times as well as its mean
$\langle T_{\rm FH} \rangle$ 
and variance 
${\rm Var}(T_{\rm FH})$.  
We also calculate the probabilities $P_{\rm BT}$ and $P_{\rm RET}$ that
the first hitting event will occur via the backtracking scenario or via the retracing scenario, respectively. 
We show that in dilute networks the dominant first hitting scenario is backtracking while in
dense networks the dominant scenario is retracing 
and calculate  
the conditional distributions  
$P(T_{\rm FH}=t| {\rm BT})$ 
and 
$P(T_{\rm FH}=t| {\rm RET})$,
for the two scenarios. 
The analytical results are 
in excellent agreement with the results 
obtained from computer simulations.
Considering the first hitting event as a termination mechanism of the RW trajectories,
these results provide useful
insight into the general problem of survival analysis and the statistics of mortality
rates when two or more termination scenarios coexist.

\end{abstract}


\noindent{\it Keywords}: 
Random network, 
random regular graph,
random walk, 
backtracking,
retracing,
first hitting time,
first intersection length.

\maketitle

\section{Introduction}

Random walk (RW) models 
\cite{Spitzer1964,Weiss1994}
are useful for the study of a large variety of
stochastic processes 
such as diffusion
\cite{Berg1993,Ibe2013},
polymer structure 
\cite{Edwards1965,Fisher1966,Degennes1979},
and random search
\cite{Evans2011,Lopez2012}.
These models were studied extensively 
in different geometries,
including
continuous space
\cite{Lawler2010b}, 
regular lattices 
\cite{Lawler2010a},
fractals 
\cite{ben-Avraham2000}
and 
random networks
\cite{Noh2004}.
In the context of complex networks
\cite{Havlin2010,Newman2010}, 
random walks 
provide useful tools for the analysis of
dynamical processes such as the
spreading of rumours, opinions and infections
\cite{Pastor-Satorras2001,Barrat2012}.

Consider an RW on a random network.
Starting at time $t=1$ from a random initial node $x_1$, 
at each time step $t \ge 2$ it hops
randomly to one of the neighbors of the previous node.
The RW thus generates a trajectory of the form
$x_1 \rightarrow x_2 \rightarrow \dots \rightarrow x_t \rightarrow \dots$,
where $x_{t}$ is the node visited at time $t$.
In some of the time steps the RW hops into nodes that
have not been visited before, while
in other time steps it hops into nodes that have
already been visited at an earlier time.
Since RWs on random networks may visit some of the nodes more than once,
the number of distinct nodes visited up to time $t$ 
cannot exceed $t$ and  
is typically smaller than $t$.
The mean number $\langle S \rangle_t$ of distinct nodes visited by an RW 
on a random network up to time $t$ was recently studied 
\cite{Debacco2015}.
It was found that 
in the infinite network limit it scales linearly with $t$, namely
$\langle S \rangle_t \simeq r t$, 
where the coefficient
$r<1$ depends on the network topology.
These scaling properties resemble those obtained
for RWs on high dimensional lattices and Cayley trees
and imply that RWs on random networks 
revisit previously visited nodes
less frequently than RWs on low dimensional lattices
\cite{Montroll1965}.
Therefore, RW models provide a highly effective framework for
search and exploration processes on random networks.

The early stages of an RW trajectory can be characterized by the distribution of
first hitting (FH) times $P(T_{\rm FH}=t)$,
where $T_{\rm FH}$ is the first time at which the RW steps into a
node which has already been visited before.
The first hitting time is also known as the first intersection length
\cite{Herrero2003,Herrero2005b}.
The first hitting event may occur either via the backtracking scenario, in which the RW 
hops back into the previous node, or by the retracing scenario, in which it hops into a node visited two
or more time steps earlier.
The first hitting event marks the transition from the first stage to the second stage
in the life cycle of an RW.
In the first stage the RW visits a new node at each time step,
while in the second stage it combines moves at which the RW visits
new nodes and moves at which it enters nodes that have already
been visited before.
The second stage 
in the life cycle of an RW on a finite network
can be characterized by the distribution of first passage (FP) times
$P(T_{\rm FP}=t)$, where $T_{\rm FP}$ is the first time at which an RW starting from a random
initial node $i$ visits a random target node $j$
\cite{Redner2001,HittingPassage}.
Finally, the cover time, at
which the RW completes visiting all the $N$ nodes in the network at least once
marks the transition from the second stage to the third stage in the life cycle of an RW
\cite{Kahn1989}.
Beyond the cover time the RW continues to revisit nodes that have already been visited before.

Another type of random walk model is the non-backtracking random walk (NBW).
At each time step the NBW steps into a random neighbor of the present node,
except for the node visited in the previous time step.
It is thus similar to the RW, except for the backtracking step which is eliminated.
The paths of NBWs have been studied on regular lattices and random graphs 
\cite{Alon2007}.
It was shown that they explore the network more efficiently than RWs. 
It was also shown
that they mix faster, namely require a shorter transient time to reach the stationary
distribution of visiting frequencies throughout the network. 

In a recent paper we presented analytical results for the distribution of 
first hitting times of RWs on Erd{\H o}s-R\'enyi (ER) networks
\cite{Tishby2017}. 
It was found that the tail distribution of first hitting times,
$P(T_{\rm FH} > t)$,
consists of a product of a 
geometric distribution due to the backtracking process
and a Rayleigh distribution due to the retracing process.
The mean and variance of the distribution of first hitting times 
were also calculated.
In this analysis we utilized a special property of RWs on ER networks,
in which the subnetwork that consists of the yet-unvisited nodes
remains an ER network at all times.
Its degree distribution remains a Poisson distribution, while its
mean degree decreases linearly with the time $t$.
This self-similarity enabled us to calculate the probability
that at time $t$ the RW will step into a yet-unvisited node
and the complementary probability that it will step into an
already visited node.
In the case of other configuration model networks, there is no
closed form expression for the time evolution of the degree distribution
of the sub-network of the yet-unvisited nodes. 
Therefore, the 
approach we used for the calculation of the distribution
of first hitting times in ER networks cannot be generalized to other
configuration model networks.
However, it turns out that in the special case of random regular graphs 
there are other simplifying
features that can be used to derive a closed form expression for
$P(T_{\rm FH} > t)$.

In another recent paper we studied the distribution of first hitting times
of NBWs on ER networks
\cite{Tishby2017b}.
Apart from the retracing scenario, NBWs on ER networks exhibit an additional 
mechanism of first hitting, referred to as the trapping scenario.
The trapping takes place when the NBW enters a leaf node of degree $k=1$,
which has no other neighbor except from the previous node.
It was found that the tail distribution of first hitting times of NBWs on ER networks
consists of a product of a 
geometric distribution due to the trapping process
and a Rayleigh distribution due to the retracing process.
The mean and variance of the distribution of first hitting times 
were also calculated.

In this paper we present analytical results for 
the distributions of first hitting times 
of random walks on random regular graphs (RRGs) of degree $c \ge 3$
and a finite size $N$. 
The first hitting event may take place either by backtracking to the previous node or by
retracing, namely stepping into a node which has been visited two or more
time steps earlier.
Using the microstructure and statistical properties of RW paths at early times,
we calculate the tail distribution of first hitting (FH) times
$P(T_{\rm FH} > t)$,
which is given by a product of a geometric distribution due to the backtracking process
and a Rayleigh distribution due to the retracing process.
We also obtain closed form expressions for 
the mean first hitting time $\langle T_{\rm FH} \rangle$ and for the variance 
${\rm Var}(T_{\rm FH})$ of the distribution of first hitting times.
The analytical results are 
found to be in excellent agreement with the results 
obtained from computer simulations.
We obtain analytical results for the probabilities $P_{\rm BT}$ and $P_{\rm RET}$ that
the first hitting event will occur via the backtracking or retracing scenarios, respectively. 
We show that in dilute networks the dominant first hitting scenario is backtracking while in
dense networks the dominant scenario is retracing. We also obtain expressions for
the conditional distributions of first hitting time, $P(T_{\rm FH}=t| {\rm BT})$ and $P(T_{\rm FH}=t| {\rm RET})$,
namely conditioned on the first hitting event occurring via the backtracking or the retracing scenario, respectively.

The paper is organized as follows.
In Sec. 2 we briefly describe the random regular graph.
In Sec. 3 we present the random walk model.
In Sec. 4 we calculate the distribution of first hitting times.
In Sec. 5 we calculate the mean first hitting time.
In Sec. 6 we calculate the variance of the distribution of first hitting times.
In Sec. 7 we analyze the interplay between the backtracking and retracing scenarios.
In Sec. 8 we compare the results presented in this paper 
for RWs on RRGs with previous results for RWs on ER networks.
The results are discussed in Sec. 9 and summarized in Sec. 10.

\section{The random regular graph}

A random network (or graph) consists of a set of $N$ nodes that
are connected by edges in a way that is determined by some
random process.
For example, in a configuration model network the degree of each node is 
drawn independently from a given degree distribution $P(k)$ and
the connections are random and uncorrelated
\cite{Newman2001}.

An important example of a configuration model network is
the Erd{\H o}s-R\'enyi (ER) network  
\cite{Erdos1959,Erdos1960,Erdos1961}.
The ER network,
denoted by
$ER(N,p)$,
consists of $N$ nodes such that each pair of nodes
is connected with probability $p$. 
The degree distribution 
of an ER network 
is a binomial distribution, $B(N,p)$.
In the limit
$N \rightarrow \infty$
and
$p \rightarrow 0$,
where the mean degree
$c=(N-1)p$
is held fixed,
it converges to a
Poisson distribution.
In the asymptotic limit 
($N \rightarrow \infty$),
the ER
network exhibits a phase transition at 
$c=1$ (a percolation transition), such that for
$c<1$
the network consists only of small clusters 
and isolated nodes, while for 
$c>1$
there is a giant cluster which includes 
a macroscopic fraction of the network, in addition
to the small clusters and isolated nodes
\cite{Molloy1995,Molloy1998}. 
At a higher value of the connectivity, namely at 
$c = \ln N$, 
there is a second transition, above which 
the entire network is included in
the giant cluster and there are 
no isolated components. 

The RRG
is a special case of a configuration 
model network, in which the degree distribution is a degenerate
distribution of the form 
$P(k)=\delta_{k,c}$, namely
all the nodes are of the same degree $c$.
Here we focus on the case of $3 \le c \le N-1$,
in which for a sufficiently large value of $N$ the RRG
consists of a single connected component.
In the infinite network limit the RRG
exhibits a tree
structure with no cycles. 
Thus, in this limit it coincides with a Bethe lattice whose coordination number is equal to $c$.
In contrast, RRGs
of a finite size exhibit a local tree-like structure,
while at larger scales there is a broad spectrum of cycle lengths
\cite{Bonneau2017}.
In that sense RRGs
differ from Cayley trees, which maintain their
tree structure by reducing the most peripheral nodes to leaf nodes of degree $1$.

A convenient way to construct an RRG
of size $N$ and degree $c$
is to prepare the $N$ nodes such that each node is 
connected to $c$ half edges or stubs
\cite{Newman2010}.
At each step of the construction, one connects a random pair of stubs that 
belong to two different nodes $i$ and $j$ 
that are not already connected,
forming an edge between them.
This procedure is repeated until all the stubs are exhausted.
The process may get stuck before completion in case that
all the remaining stubs belong to the
same node or to pairs of nodes that are already connected.
In such case one needs to perform some random reconnections
in order to complete the construction.

RRGs provide a useful benchmark for the study of dynamical processes
on regular lattices. This is due to the fact that for any regular lattice structure in any
space dimension, one can construct an RRG
whose degree $c$
is equal to the coordination number $z$ of the regular lattice.
For example, a simple cubic lattice in a $d$ dimensional space corresponds to 
an RRG
with degree $c=2d$.
Some special cases in two dimensions include the honeycomb lattice that 
corresponds to an RRG
with $c=3$ and the triangular lattice
that corresponds to an RRG
with $c=6$.
However, all the other structural properties of RRGs 
are completely different from those of the corresponding regular lattice.
Therefore, comparing the behavior of dynamical processes taking place on regular lattices 
and on the corresponding RRGs,
one can distinguish between the properties
that depend on the coordination number and those that depend on other structural properties
such as the lengths of short cycles.
A famous example is the Bethe-Peierls approximation of the Ising model
\cite{Pathria2011}, which is actually an exact result on an RRG
with the same coordination number as a $d$-dimensional lattice.
In $2$ dimensions
the critical exponents are wrongly identified as the mean-field ones,
but the actual predictions for the critical temperature and the heat capacity
are highly insightful.
In dimensions higher than $4$, even the critical exponents are correct.

\section{The random walk model}

Consider an RW on an RRG
of degree $c \ge 3$ and size $N$.
At each time step the RW hops from its current node to one of its neighbors,
such that the probability of hopping to each neighbor is $1/c$.
For sufficiently large $N$ the RRG
consists of a single connected
component, thus an RW starting from any initial node
can reach any other node in the network.
In the long time limit $t \gg N$ the RW visits all the nodes with the same frequency,
namely on average each node is visited once every $N$ steps.
However, over shorter periods of time there may be large fluctuations such that
some nodes may be visited several times
in a given time interval
while other nodes are not visited at all.

In some of the time steps an RW may visit nodes that have not been visited before
while in other time steps it may revisit nodes that have already been visited before. 
For example, at each time step $t \ge 3$ the RW may backtrack into the previous node with 
probability of $1/c$.
In the infinite network limit the RRG
exhibits a tree structure.
Therefore, in this limit the backtracking mechanism is the only way in which an RW
may hop from a newly visited node to a node that has already been visited before.
However, in finite networks the RW may utilize the cycles to retrace its path 
and hop into nodes that have already been visited two or more time steps earlier.
Once the RW stepped into a node that has been visited before,
it may continue to hop back and forth along
the path of its previously visited nodes, 
until it eventually leaves the path and enters a newly visited node.

\begin{figure}
\centerline{
\includegraphics[width=5.5cm]{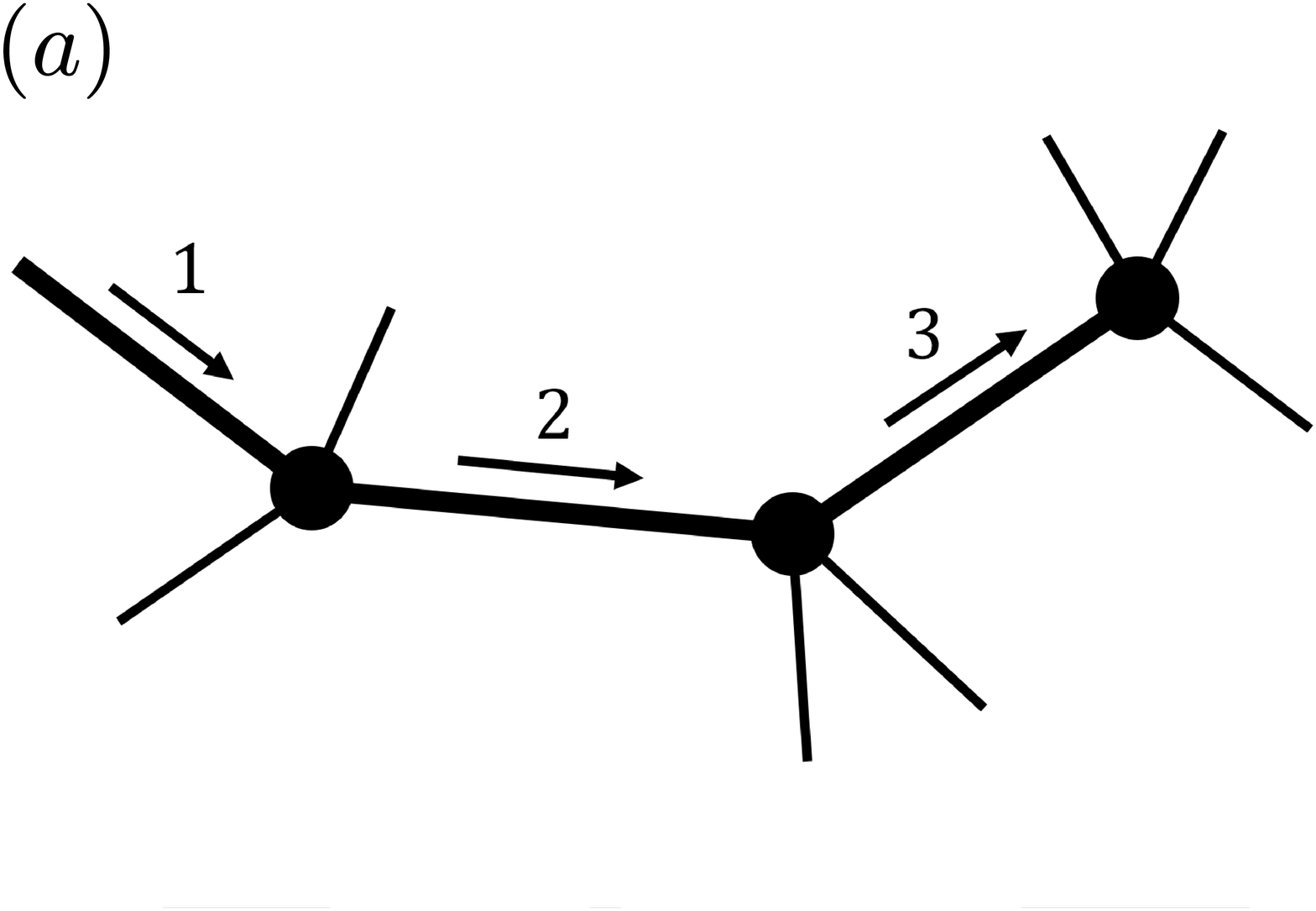}
}
\centerline{
\includegraphics[width=5.5cm]{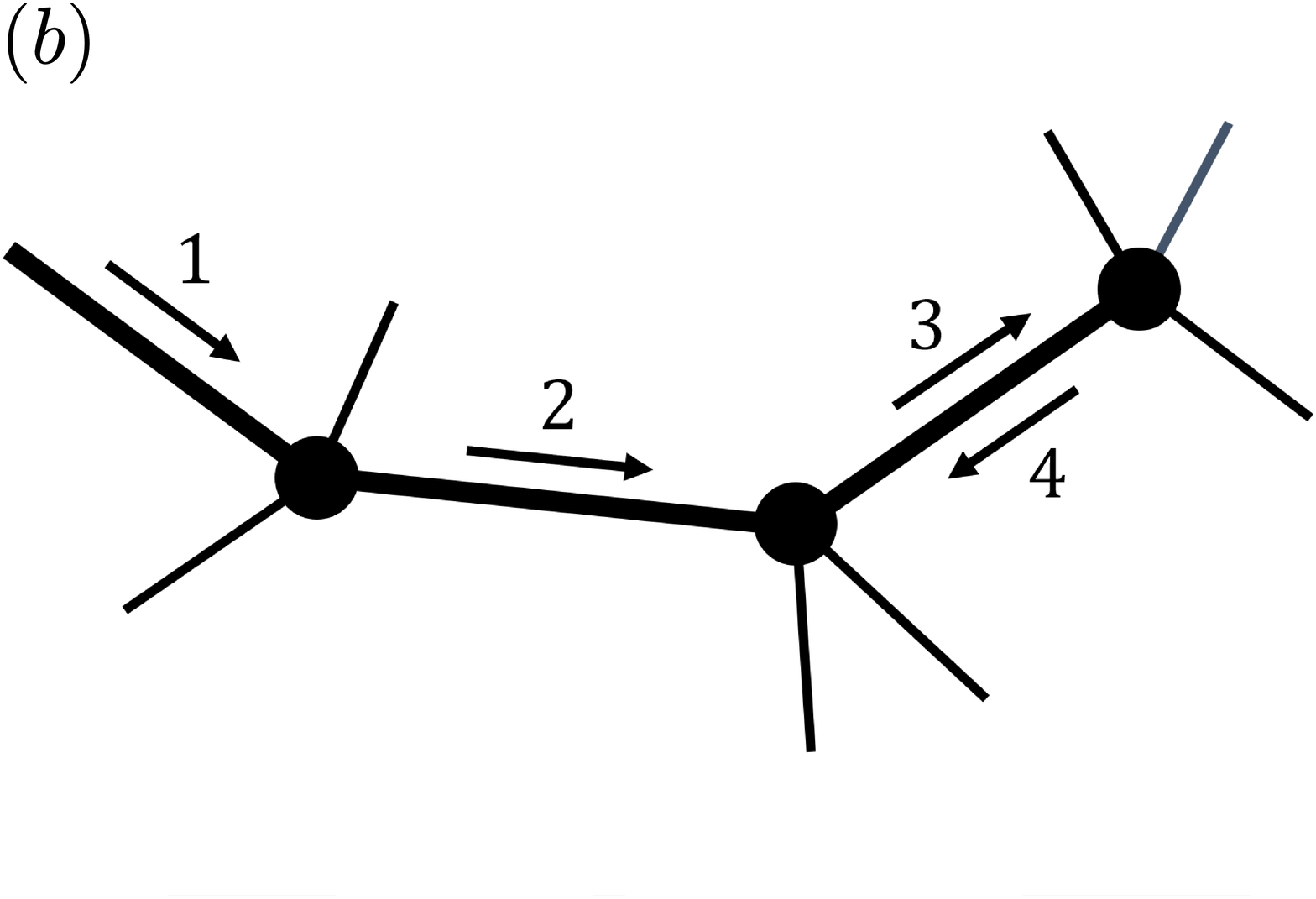}
}
\centerline{
\includegraphics[width=5.5cm]{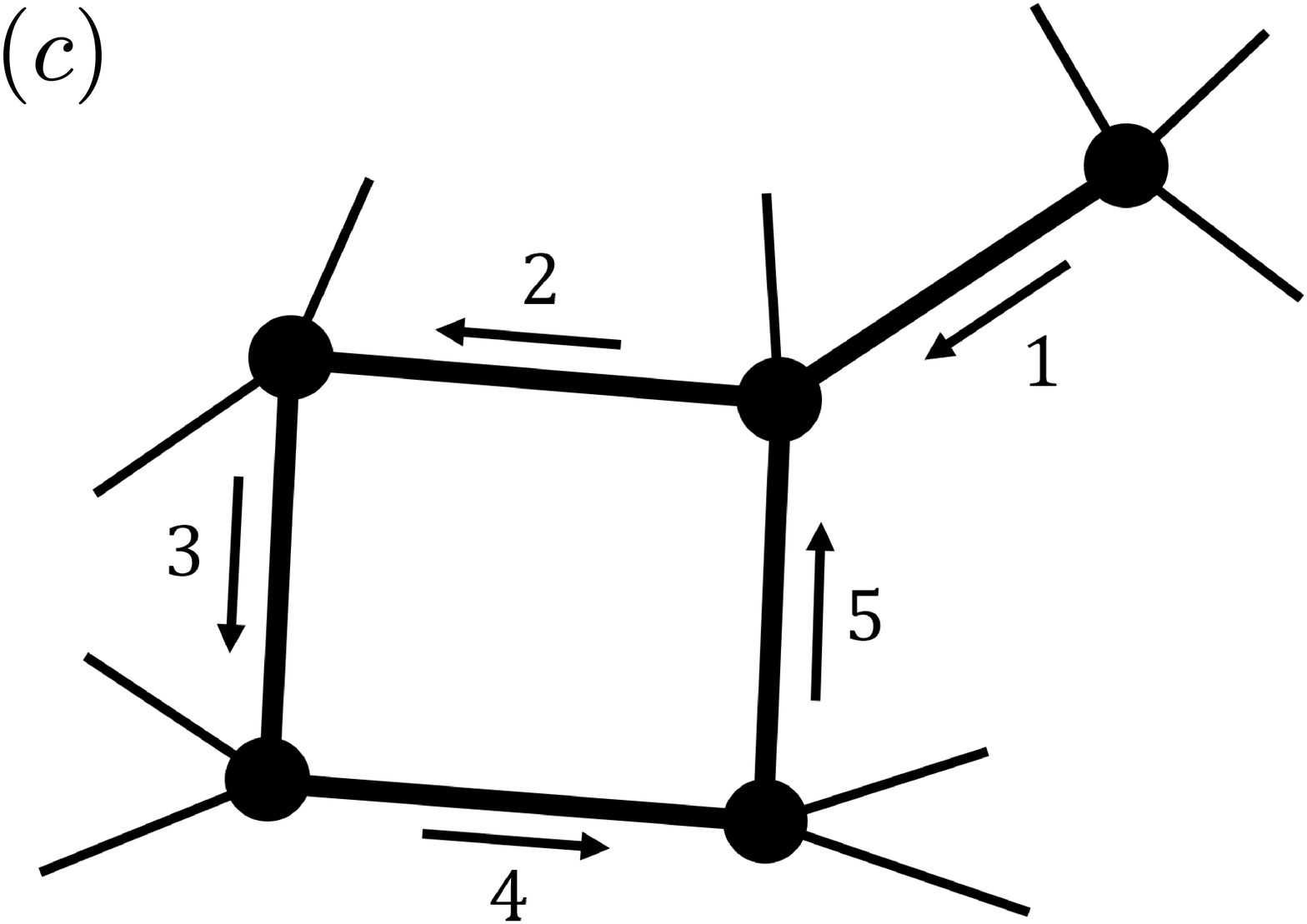}
}
\caption{
Schematic illustrations of possible events taking place along the path 
of an RW on an RRG:
(a) a path segment in which at each time step the RW enters a node
that has not been visited before; (b) a path segment that includes a backtracking step
into the previous node (step no. 4);
(c) a path that includes a retracing step (step no. 5) in which the RW hops into a node that was
visited a few time steps earlier. Retracing steps are not possible in the infinite
network limit and take place only in finite networks, which include cycles. 
Note that in this illustration the RRG
is of degree $c=4$.
}
\label{fig:1}
\end{figure}

In Fig. \ref{fig:1} we present a
schematic illustration of the backtracking and the retracing
events which may take place along the path 
of an RW on an RRG.
While backtracking is independent on the network size,
retracing takes place only in finite networks
and is not possible in the infinite network limit, in which the RRG
exhibits a tree structure.

\section{The distribution of first hitting times}

Consider an RW on an RRG 
of a finite size $N$ and degree $c \ge 3$.
Starting from a random node at time $t=1$,
the RW hops randomly between nearest neighbor nodes.
At early times $t \ll N$ all the nodes it enters are likely to be visited for the first time.
The first time at which the RW enters a node that has already been visited before
is called the first hitting time.
The first hitting process may take place either by
backtracking (BT) or by retracing (RET).
In the backtracking scenario 
the RW moves back into the previous node,
while in the retracing scenario it hops into a node that has already 
been visited two or more time steps earlier.
Below we calculate the distribution of first hitting times.
In case that the RW has not returned to any previously visited node 
up to time $t-1$, the first hitting time satisfies the condition 
$T_{\rm FH}>t-1$.
Given that
$T_{\rm FH}>t-1$,
the probability that the first hitting event will not take place in the next time step is
expressed by the conditional probability
$P(T_{\rm FH} > t |T_{\rm FH} > t - 1)$. 
This conditional probability can be expressed as a product of the form

\begin{equation}
P(T_{\rm FH}>t|T_{\rm FH}>t-1) =
P_t( \lnot{\rm  RET} | \lnot {\rm BT} )
P_t(\lnot {\rm BT}),
\label{eq:PFHbr}
\end{equation}

\noindent 
where
$P_t(\lnot {\rm BT})$
is the probability that the RW will not backtrack to the previous node
at time $t$.
Given that the RW has not backtracked at time $t$, the conditional
probability
$P_t( \lnot {\rm  RET} | \lnot {\rm BT} )$
is the probability that it will also not 
retrace its path at time $t$, namely, that it will not hop into
a node that has already been visited two or more time steps earlier.

Since all the nodes in the RRG
are of degree $c$,
at each time step (apart from $t=1$ and $2$)
the probability of backtracking into the previous node
is $1/c$. Therefore, the probability that a backtracking step will not occur
at any given time $t \ge 3$ is

\begin{equation}
P_t(\lnot {\rm BT})
=
1 - \frac{1}{c}.
\label{eq:PFHb}
\end{equation}

\noindent
Provided that the RW has not backtracked at time $t$,
we will now evaluate the probability
$P_t( \lnot {\rm  RET} | \lnot {\rm BT} )$
that it will also not retrace its path.  
Apart from the current node at time $t-1$ and the previous node 
(visited at time $t-2$), there are $N-2$ other nodes
in the network,
of which $t-3$ have already been visited by the RW.

Since the network is undirected, an edge connecting a pair 
of nodes $i$ and $j$ can be considered as a combination of
two links, one from $i$ to $j$ and the other from $j$ to $i$.
Thus, a node of degree $c$ is connected to $c$ incoming links and $c$
outgoing links.
The number of yet-unvisited nodes at time $t-1$ is 
$N-t+1$. Each one of these nodes is of degree $c$.

Since at each time step the RW enters a node via one edge and leaves it
via another edge, up to time $t-1$ it exhausts $t-2$ edges along the path. 
Thus, each one of the nodes visited in the first $t-3$ time steps exhibits
only $c-2$ links that can be used by the RW to enter the node via the retracing mechanism
at time $t$
(apart from the initial node that can be reached via $c-1$ edges).
As a result, the number of incoming links that may lead the RW to an already visited
node via the retracing mechanism at time $t$ is 

\begin{equation}
L_{\rm v} = (c-2)(t-3)+1.
\end{equation}

\noindent
The number of incoming links that may lead the RW to a yet-unvisited node 
at time $t$ is

\begin{equation}
L_{\rm u} =(N - t + 1)c.
\end{equation}

\noindent
Summing up $L_{\rm v}$ and $L_{\rm u}$ we obtain the
total number of incoming links that may lead the RW 
at time $t$
to either a visited node
(by retracing) or to a
yet-unvisited node. It
is given by

\begin{equation}
L_{\rm T} = (N-2)c - 2(t-3) + 1.
\end{equation}

\noindent
Given that the possibility of backtracking into the previous node was eliminated, 
the RW selects randomly
one of the $c-1$ other neighbors of the current node.
The probability that any one of these neighbors has already been 
visited before is thus given by $L_{\rm v}/L_{\rm T}$.
Therefore, the probability of retracing at time $t$
under the condition of no-backtracking
is given by

\begin{equation}
P_t(   {\rm  RET} | \lnot {\rm BT} )
=
\frac{ L_{\rm v} }{ L_{\rm T} }
=
\frac{(c-2)(t-3)+1}{  (N-2)c - 2(t-3) + 1}.
\label{eq:PFHra2}
\end{equation}

\begin{figure}
\centerline{
\includegraphics[width=9cm]{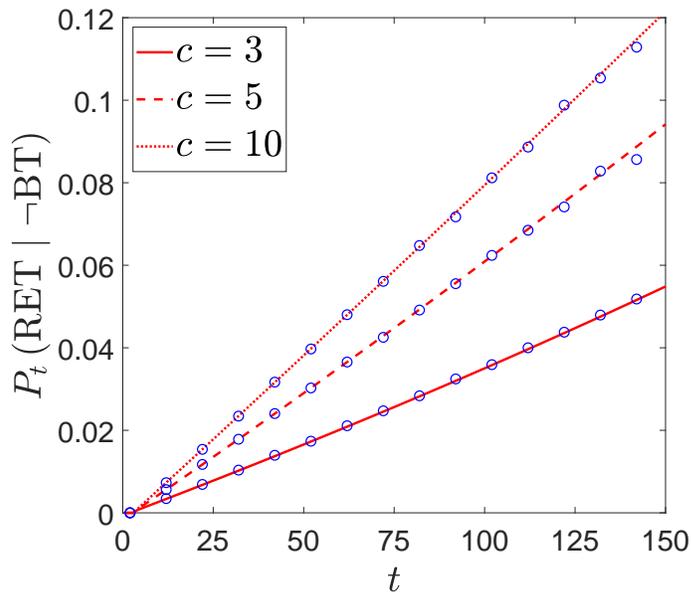}
}
\caption{
Analytical results for the probability
$P_t({\rm RET} | \lnot {\rm BT})$  
that an RW will retrace its path at time $t$
under the condition that it has not backtracked
its path at time $t$,
on an RRG 
of size $N=1000$ and 
$c=3$ (solid line), $5$ (dashed line) and $10$ (dotted line).
The analytical results, obtained from Eq. (\ref{eq:PFHra2}), are
in excellent agreement with the results obtained from
computer simulations (circles).
}
\label{fig:2}
\end{figure}

\noindent
In Fig. \ref{fig:2} we present
analytical results,
obtained from Eq. (\ref{eq:PFHra2}),
for the probability
$P_t({\rm RET} | \lnot {\rm BT})$  
that an RW on an RRG
will retrace its path
for the first time
at time $t$
under the condition that it has not backtracked
its path at time $t$ (or earlier).
The RRGs 
used in Fig. \ref{fig:2} are
of size $N=1000$, 
where the nodes are of degree
$c=3$ (solid line), $5$ (dashed line) and $10$ (dotted line).
The analytical results are in excellent agreement with the results obtained
from computer simulations (circles).
These results are qualitatively different from the corresponding
results for RWs on ER networks, which are given by
\cite{Tishby2017}

\begin{equation}
P_t^{\rm ER}(   {\rm  RET} | \lnot {\rm BT} )
=
\frac{c(t-3)}{(c+1)(N-1)},
\label{eq:PFHra2b}
\end{equation}

\noindent
where $c$ is the mean degree.

Using Eq. (\ref{eq:PFHra2}),
the probability that the RW will not retrace its path at time $t \ge 3$ 
is given by

\begin{equation}
P_t( \lnot {\rm  RET} | \lnot {\rm BT} )
=
1 -   \frac{(c-2)(t-3)+1}{(N-2)c-2(t-3)+1}.
\label{eq:PFHr}
\end{equation}

\noindent
Inserting the probabilities
$P_t(\lnot {\rm BT})$
and
$P_t( \lnot {\rm  RET} | \lnot {\rm BT} )$
from Eqs. (\ref{eq:PFHb}) and (\ref{eq:PFHr}),
respectively, into Eq. (\ref{eq:PFHbr}),
we obtain

\begin{equation}
P(T_{\rm FH}>t|T_{\rm FH}>t-1) =
\left( 1 - \frac{1}{c} \right)   \left[ 1  -   \frac{(c-2)(t-3)+1}{(N-2)c-2(t-3)+1}  \right].
\label{eq:PFHbr2b}
\end{equation}

\noindent
Focusing on large networks, where $N \gg 1$,
and assuming that $t \ll Nc$,
the probability 
$P_t( \lnot {\rm  RET} | \lnot {\rm BT} )$
can be written as a series expansion in powers
of $(t-3)/(Nc)$. It is given by

\begin{equation}
P_t( \lnot  {\rm  RET} | \lnot {\rm BT} )
=
1 -
\frac{(c-2)(t-3) }{Nc }
-
\frac{2(c-2)(t-3)^2}{(Nc)^2}
+
\mathcal{O} \left[ \left( \frac{t-3}{Nc}  \right)^3  \right].
\label{eq:PFHra7}
\end{equation}

\noindent
Using the expansion $\ln (1-x) = - x - x^2/2 + \mathcal{O}(x^3)$
for $x \ll 1$, we obtain

\begin{eqnarray}
\ln \left[ P_t( \lnot {\rm  RET} | \lnot {\rm BT} ) \right]
&\simeq&
  -    \frac{(c-2)(t-3) }{Nc }
-
\frac{2(c-2)(t-3)^2}{(Nc)^2}
\nonumber \\
&-&   \frac{1}{2}  \left[  \frac{(c-2)(t-3) }{Nc }
+
\frac{2(c-2)(t-3)^2}{(Nc)^2}      \right]^2.
\label{eq:PFHrb}
\end{eqnarray}

\noindent
Rearranging terms up to second order in $(t-3)/(Nc)$, we obtain

\begin{equation}
P_t( \lnot {\rm  RET} | \lnot {\rm BT} )
\simeq
\exp \left[ 
- \frac{(c-2)(t-3)}{Nc} 
- \frac{(c-2)(c+2)(t-3)^2}{2(Nc)^2} 
 \right].
\label{eq:PFHrexp}
\end{equation}

\noindent
As a result, Eq. (\ref{eq:PFHbr2b}) is replaced by

\begin{equation}
P(T_{\rm FH}>t|T_{\rm FH}>t-1) \simeq
\left( 1 - \frac{1}{c} \right) \exp \left[   
-   \frac{(c-2)(t-3) }{Nc }
-      \frac{(c-2) (c+2) (t-3)^2 }{2 (Nc)^2}
  \right],
\label{eq:PFHbr2}
\end{equation}

\noindent
where $t \ge 3$.

The probability $P(T_{\rm FH}>t)$ 
that the first hitting event will not take place during
the first $t$ time steps is given by the product

\begin{equation}
P(T_{\rm FH}>t) = 
\prod_{t'=3}^{t}
P(T_{\rm FH}>t'|T_{\rm FH}>t'-1).
\label{eq:prod}
\end{equation}

\noindent
Inserting $P(T_{\rm FH}>t'|T_{\rm FH}>t'-1)$
from Eq. (\ref{eq:PFHbr2}) into Eq. (\ref{eq:prod}),
we obtain the tail distribution of first hitting times

\begin{equation}
P( T_{\rm FH} > t ) = 
P_{\rm BT}( T_{\rm FH}>t)
P_{\rm RET}( T_{\rm FH}>t),
\label{eq:PTFHt1}
\end{equation}

\noindent
where

\begin{equation}
P_{\rm BT}( T_{\rm FH} > t ) = 
\left\{
\begin{array}{ll}
1 & \ \ \ \ \ \ \  t=1,2 \\
\left(1-\frac{1}{c}\right)^{t-2}   & \ \ \ \ \ \ \  t \ge 3,
\end{array}
\right.  
\label{eq:Pbt7}
\end{equation}

\noindent
and

\begin{equation}
P_{\rm RET}( T_{\rm FH} > t ) \simeq 
\left\{
\begin{array}{ll}
1 & \ \ \ \ \ \ \  1 \le t \le 3 \\
\prod_{t'=3}^{t} \exp \left[-     \frac{(c-2)(t'-3) }{Nc} 
-    \frac{(c-2) (c+2) (t'-3)^2 }{2 (Nc)^2}  \right]   & \ \ \ \ \ \ \  t \ge 4. 
\end{array}
\right.  
\label{eq:Pret7}
\end{equation}

\noindent
Note that the contribution of the backtracking process, given by Eq. (\ref{eq:Pbt7}),
depends only on the mean degree $c$, while the contribution of the retracing process,
given by Eq. (\ref{eq:Pret7}) depends on the mean degree $c$ and on the network size $N$.
In the infinite network limit the retracing scenario becomes irrelevant and Eq. (\ref{eq:PTFHt1}) 
is simplified to

\begin{equation}
P( T_{\rm FH} > t | N \rightarrow \infty ) = 
\left\{
\begin{array}{ll}
1 & \ \ \ \ \ \ \  t=1,2 \\
\left(1-\frac{1}{c}\right)^{t-2}  & \ \ \ \ \ \ \  t \ge 3. 
\end{array}
\right.  
\label{eq:PTFHt2}
\end{equation}

\noindent
Taking the logarithm of $P_{\rm RET}(T_{\rm FH}>t)$,
as expressed in Eq. (\ref{eq:Pret7})  
for $t \ge 4$, we obtain

\begin{equation}
\ln [ P_{\rm RET}( T_{\rm FH} > t ) ] \simeq 
     -   \sum_{t'=3}^{t}  
   \frac{(  c-2 )(t'-3) }{Nc } - \sum_{t'=3}^{t}  \frac{(c-2)(c+2)(t'-3)^2 }{2 (Nc)^2}   .
\label{eq:PTFHt0}
\end{equation}

\noindent
Carrying out the summations in Eq. (\ref{eq:PTFHt0}), we obtain

\begin{equation}
\ln \left[ P_{\rm RET}( T_{\rm FH} > t ) \right] \simeq 
 - \frac{(c-2)(t-2)(t-3)}{2Nc} 
- \frac{(c-2)(c+2)(t-2)(t-3)(t-5/2)}{6 (Nc)^2} .
\label{eq:PTFHt3}
\end{equation}

\noindent
Inserting 
$P_{\rm BT}( T_{\rm FH} > t )$
from Eq. (\ref{eq:Pbt7})
and
$P_{\rm RET}(T_{\rm FH}>t)$
from Eq. (\ref{eq:PTFHt3})
into Eq. (\ref{eq:PTFHt1}), we obtain

\begin{equation}
P ( T_{\rm FH}>t )
\simeq
\left\{
\begin{array}{ll}
1 & \ \ \ \ \ \ \  t=1,2 \\
\exp \left[-\frac{  (t-2)  ( t-3 )}{2  \alpha^2}
- \frac{(t-2)(t-3)(t-5/2)}{6 \left( \frac{c-2}{c+2} \right) \alpha^4}
- 
\beta
(t-2) \right]   & \ \ \ \ \ \ \  t \ge 3, 
\end{array}
\right.  
\label{eq:FH_tail0}
\end{equation}

\noindent
where the parameters
$\alpha$ and $\beta$  
are given by

\begin{equation}
\alpha = \sqrt{ \left( \frac{ c  }{c-2} \right) N },
\label{eq:FHTa}
\end{equation}

\noindent
and

\begin{equation}
\beta = \ln \left( \frac{c}{c-1} \right).
\label{eq:FHTb}
\end{equation}

\noindent
Using a Taylor expansion of the second term in the exponent in
Eq. (\ref{eq:FH_tail0}), we obtain

\begin{equation}
P ( T_{\rm FH}>t )
\simeq
\left\{
\begin{array}{ll}
1 & \ \ \ \ \ \ \  t=1,2 \\
\left[ 1 - \frac{(t-2)(t-3)(t-5/2)}{6 \left( \frac{c-2}{c+2} \right) \alpha^4} \right]
\exp \left[-\frac{  (t-2)  ( t-3 )}{2  \alpha^2}
- 
\beta
(t-2) \right]   & \ \ \ \ \ \ \  t \ge 3, 
\end{array}
\right.  
\label{eq:FH_tail7}
\end{equation}

\noindent
where the first term in the second line is the leading term and the second term
is a correction term, which is small for sufficiently small values of $t$.

It turns out that for large networks the second term in the second line
of Eq. (\ref{eq:FH_tail7})
is extremely small and its effect on the tail distribution
$P(T_{\rm FH}>t)$ 
can be neglected in most cases.
The only case in which it was found to make a small but noticeable difference is
in the calculation of the second moment $\langle T_{\rm FH}^2 \rangle$.
For the sake of consistency, we use Eq. (\ref{eq:FH_tail7}) in the calculations
of both the mean first hitting time $\langle T_{\rm FH} \rangle$
and the second moment $\langle T_{\rm FH}^2 \rangle$.
In all the other calculations presented in this paper we use a simplified form
of the tail distribution, which is given by

\begin{equation}
P ( T_{\rm FH}>t )
\simeq
\left\{
\begin{array}{ll}
1 & \ \ \ \ \ \ \  t=1,2 \\
\exp \left[-\frac{  (t-2)  ( t-3 )}{2  \alpha^2}
- 
\beta
(t-2) \right]   & \ \ \ \ \ \ \  t \ge 3, 
\end{array}
\right.  
\label{eq:FH_tail}
\end{equation}

\noindent
The tail distribution
$P ( T_{\rm FH}>t )$,
given by Eq. (\ref{eq:FH_tail}),
is thus a product of 
a geometric distribution, associated with the backtracking process and
a Rayleigh distribution, associated with the retracing process 
\cite{Papoulis2002}.

\begin{figure}
\centerline{
\includegraphics[width=14cm]{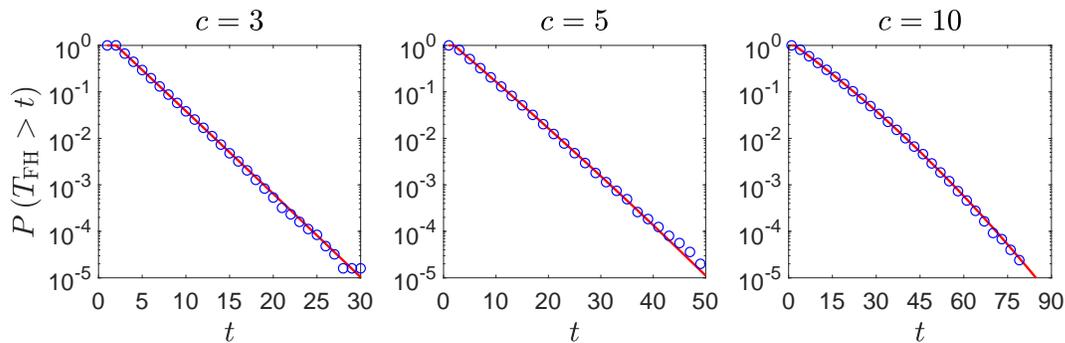}
}
\caption{
Analytical results for the tail distributions 
$P(T_{\rm FH} > t)$  
of first hitting times (solid lines)
of RWs on RRGs
of size $N=1000$, 
in which the nodes are of degree
$c=3$ (left), $5$ (middle) and $10$ (right).
The analytical results, obtained from Eq.
(\ref{eq:FH_tail}) 
are in excellent agreement with the results obtained
from computer simulations (circles).
}
\label{fig:3}
\end{figure}

In Fig. \ref{fig:3} 
we present 
(on a semi-logarithmic scale)
analytical results for
the tail distributions 
$P( T_{\rm FH} > t)$
of first hitting times (solid lines)
of RWs on RRGs
of size $N=1000$ 
in which the nodes are of degree 
$c=3$ (left), $5$ (middle) and $10$ (right).
The analytical results, obtained from Eq.
(\ref{eq:FH_tail}) 
are in excellent agreement with the results obtained
from computer simulations (circles).
It is found that as the degree $c$ is increased
the first hitting event tends to occur at a later time.
This can be attributed to the fact that the probability
of backtracking decreases as $c$ is increased.

The excellent agreement observed in Fig. \ref{fig:3} between the analytical
results and the results obtained from computer simulations indicates that
Eq. (\ref{eq:FH_tail})
is valid for a broad range of parameters.
However, 
Eq. (\ref{eq:FH_tail})
combines the backtracking and retracing mechanisms.
Thus, for parameters in which the backtracking mechanism is dominant,
Fig. \ref{fig:3} is not sufficient to establish the validity of the first term in
Eq. (\ref{eq:FH_tail}),
associated with the retracing scenario.
This is due to the fact that in this regime the retracing process is effectively
screened by the backtracking process.

In order to establish the validity of the retracing term in 
Eq. (\ref{eq:FH_tail})
we consider the distribution of first hitting times of NBWs
on RRGs.
The suppression of the backtracking process in NBWs leaves
the retracing scenario as the only possible scenario of first hitting
in these systems.
This is unlike the case of NBWs on ER networks, in which the
trapping scenario emerges once backtracking is suppressed.
This contrast is due to the fact that RRGs
do not include leaf nodes of degree $k=1$, in which an NBW may
become trapped.
The suppression of the backtracking mechanism implies that the
distribution of first hitting times of NBWs on RRGs
is expressed by the right hand side of Eq. (\ref{eq:FH_tail}),
where $\alpha$ is given by Eq. (\ref{eq:FHTa})
and $\beta=0$, namely

\begin{equation}
P_{\rm NBW} (  T_{\rm FH}>t  )
\simeq
\left\{
\begin{array}{ll}
1 & \ \ \ \ \ \ \  t=1,2 \\
\exp \left[-\frac{  (t-2)  ( t-3 )}{2  \alpha^2}
\right]   & \ \ \ \ \ \ \  t \ge 3. 
\end{array}
\right.  
\label{eq:FH_tail2}
\end{equation}

\begin{figure}
\centerline{
\includegraphics[width=14cm]{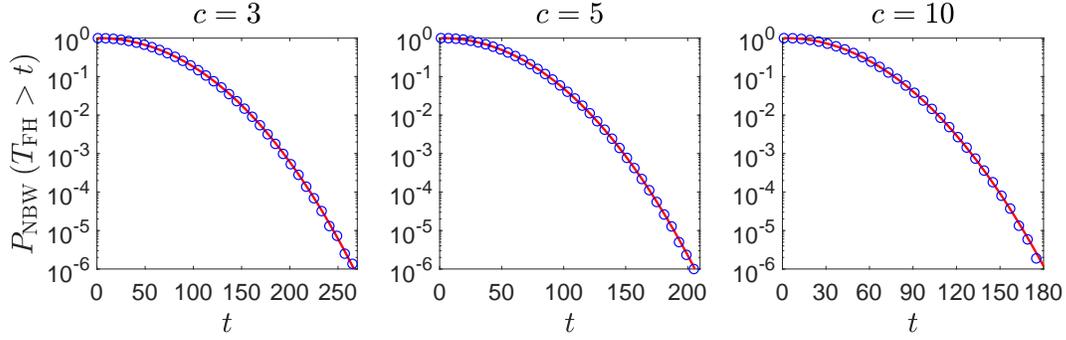}
}
\caption{
Analytical results for the tail distributions 
$P_{\rm NBW}(T_{\rm FH} > t)$  
of first hitting times (solid lines)
of NBWs on RRGs
of size $N=1000$, 
in which the nodes are of degree
$c=3$ (left), $5$ (middle) and $10$ (right).
The analytical results, obtained from Eq.
(\ref{eq:FH_tail2}) 
are in excellent agreement with the results obtained
from computer simulations (circles).
}
\label{fig:4}
\end{figure}

In Fig. \ref{fig:4} 
we present 
(on a semi-logarithmic scale)
analytical results for
the tail distributions 
$P_{\rm NBW}( T_{\rm FH} > t)$
of first hitting times (solid lines)
of NBWs on RRGs
of size $N=1000$ 
in which the nodes are of degree 
$c=3$ (left), $5$ (middle) and $10$ (right).
The analytical results, obtained from Eq.
(\ref{eq:FH_tail2}) 
are in excellent agreement with the results obtained
from computer simulations (circles).

\section{The mean first hitting time}

The moments 
$\langle T_{\rm FH}^r \rangle$, $r=1,2,\dots$,
of the distribution of first hitting times 
of RWs on RRGs
can be obtained 
from the tail-sum formula
\cite{Pitman1993}

\begin{equation}
\langle T_{\rm FH}^r \rangle = \sum_{t=0}^{N} 
[ (t+1)^r - t^r ] P(T_{\rm FH}>t).
\label{eq:FH_r}
\end{equation} 

\noindent
In particular, the
mean first hitting time 
$\langle T_{\rm FH} \rangle$ 
can be obtained by inserting $r=1$ in Eq. (\ref{eq:FH_r}),
which yields

\begin{equation}
\langle T_{\rm FH} \rangle = \sum_{t=0}^{N} P(T_{\rm FH}>t).
\label{eq:FH_sum}
\end{equation} 

\noindent
Inserting $P(T_{\rm FH}>t)$ 
from Eq. (\ref{eq:FH_tail7}) into Eq. (\ref{eq:FH_sum}),
shifting the summation index from $t$ to $t-1$, and
and replacing the sum in Eq. (\ref{eq:FH_sum}) by a sum of two integrals,
using the formulation of a middle Riemann sum,
we obtain

\begin{equation}
\langle T_{\rm FH}  \rangle 
\simeq I_1 + I_2,
\label{eq:<TFH>}
\end{equation}

\noindent
where

\begin{equation}
I_1 =
3 + 
e^{-\beta}
+
\int\limits_{3/2}^{N-3/2}
\exp \left[ -\frac{t(t-1)}{2 \alpha^2} - 
\beta t \right] dt,
\label{eq:FH_int}
\end{equation}

\noindent
and

\begin{equation}
I_2 =
- \int\limits_{3/2}^{N-3/2}
\frac{t(t-1)(t-1/2)}{6 \left( \frac{c-2}{c+2} \right) \alpha^4}
\exp \left[ -\frac{t(t-1)}{2 \alpha^2} - 
\beta t \right] dt.
\label{eq:FH_int2}
\end{equation}

\noindent
Carrying out the integration in Eq. (\ref{eq:FH_int}), we obtain  

\begin{equation}
I_1
\simeq
3 + e^{-\beta} + \sqrt{\frac{\pi}{2}}\alpha 
\exp \left[  
\frac{ \left( 2 \alpha^2 \beta - 1 \right)^2 }{8 \alpha^2}
\right]
\left\{ {\rm erf} \left[ \frac{\alpha (\beta+1 - 2/c) }{\sqrt{2}} \right]
- {\rm erf}\left(\frac{\alpha^2 \beta + 1}{\sqrt{2} \alpha}\right)\right\},
\label{eq:TFH1}
\end{equation}

\noindent
where ${\rm erf}(x)$ is the error function, 
also called Gauss error function
\cite{Olver2010}.
The error function ${\rm erf}(x)$ 
is a monotonically increasing function, defined for $-\infty < x < \infty$.
It is an odd function, namely ${\rm erf}(-x) = - {\rm erf}(x)$, and 
exhibits a sigmoidal shape. For 
$|x| \ll 1$ 
it can be
approximated by 
${\rm erf}(x) \simeq 2x/\sqrt{\pi}$ 
while for 
$|x| > 1$ 
it quickly converges to 
${\rm erf}(x) \rightarrow {\rm sign}(x)$.
The parameter $\beta$ 
obtains its largest value at $c=3$, where $\beta= \ln(3/2)$,
and decreases monotonically as $c$ is increased.
Thus, for any value of $c$ it satisfies $\beta < 1$.
In the limit of $c \gg 1$, it can be approximated by

\begin{equation}
\beta =  \frac{1}{c}   +   \frac{1}{2c^2}  
+ \mathcal{O} \left( \frac{1}{c^3} \right).
\end{equation}

\noindent
Since the degree $c$ satisfies $c<N-1$, the parameter $\beta$ is bounded from below by
$\beta > 1/(N-1)$. In the limit of complete graph it approaches 
this lower bound, namely $\beta \rightarrow 1/(N-1)$.
The product $\alpha \beta$ thus takes values in the range
$ 1/\sqrt{N-2}  <  \alpha \beta \le  \ln(3/2) \sqrt{N-2}$.

Eq. (\ref{eq:TFH1}) can be written in the form

\begin{eqnarray}
I_1
\simeq 
3 + e^{-\beta} 
&+& \sqrt{\frac{\pi}{2}}\alpha 
\exp \left( {\frac{- \beta}{2}} \right)
\exp \left( {\frac{1}{8 \alpha^2}} \right)
\exp \left[
\left( \frac{ \alpha \beta }{\sqrt{2} } \right)^2
\right] 
\times
\nonumber \\
& &
\left[ {\rm erf} \left(  \frac{ \alpha (1-2/c)  }{ \sqrt{2}}  +  \frac{\alpha \beta  }{\sqrt{2}}  \right)
- {\rm erf}\left( \frac{ \alpha  \beta  }{\sqrt{2}  }   + \frac{ 1  }{\sqrt{2}  \alpha }  \right)\right].
\label{eq:TFH1b}
\end{eqnarray}

\noindent
In the large network limit, where $N \gg 1$, the term
$\exp[1/(8 \alpha^2)]$ satisfies
$\exp[1/(8 \alpha^2)] \simeq 1 + \mathcal{O}(1/N)$ 
and thus it can be neglected.
Similarly, 
in the large network limit
the argument of the first ${\rm erf}$ function 
in Eq.
(\ref{eq:TFH1})
is always very large.
Therefore, one can safely set the first ${\rm erf}$ function to be equal to $1$.
The second ${\rm erf}$ function can be approximated by

\begin{equation}
{\rm erf}\left( \frac{ \alpha  \beta  }{\sqrt{2}  }   + \frac{ 1  }{\sqrt{2}  \alpha }  \right) 
\simeq
{\rm erf}\left( \frac{ \alpha  \beta  }{\sqrt{2}  }      \right) 
+ \frac{\sqrt{2} }{ \sqrt{  \pi  } \alpha }
\exp \left[
- \left( \frac{ \alpha \beta }{\sqrt{2} } \right)^2 \right]
\end{equation}

\noindent
Making these approximations, we obtain

\begin{equation}
I_1
\simeq
3 + e^{-\beta}  - e^{-\beta/2}
+ \sqrt{\frac{\pi}{2}}\alpha 
e^{\frac{- \beta}{2} } 
\exp \left[    \left( \frac{\alpha  \beta}{\sqrt{2}} \right)^2   \right]
\left[ 1 - {\rm erf}\left(\frac{\alpha\beta}{\sqrt{2}}\right)
\right].
\label{eq:T_FHT1}
\end{equation}

Carrying out the integration in Eq. (\ref{eq:FH_int2}) and taking the large network limit, 
we obtain

\begin{eqnarray}
I_2 &\simeq&
\frac{1}{6} \sqrt{\frac{\pi }{2}} 
\left( \frac{ c+2 }{c-2} \right)
  \alpha  \beta  
   \left(\alpha ^2 \beta ^2+3\right) 
e^{\frac{\left(1-2 \alpha ^2 \beta \right)^2}{8 \alpha ^2}}
\left[1-\text{erf}\left(\frac{\alpha ^2 \beta +1}{\sqrt{2} \alpha }\right) \right]
\nonumber \\
&-& \frac{1}{6} \left( \frac{ c+2 }{c-2} \right)
   \left(\alpha ^2 \beta ^2-\beta +2\right) 
e^{-\frac{3}{8 \alpha ^2}-\frac{3 \beta }{2}}.
\label{eq:T_FHT2}
\end{eqnarray}

In the dilute-network limit, where
$3 \le c \ll 1.63 \sqrt{N}$, the argument of the error function 
in Eq. (\ref{eq:T_FHT1})
satisfies the
condition $\alpha \beta / \sqrt{2} \gg 1$.
In this limit one can use the approximation
\cite{Olver2010}

\begin{equation}
1 - {\rm erf}(x) \simeq \frac{ e^{-x^2} }{ x \sqrt{\pi} },
\end{equation}

\noindent
which is the first term in the asymptotic expansion of the
complementary error function, and obtain

\begin{equation}
\langle T_{\rm FH} \rangle \simeq 
c + 2 + \mathcal{O} \left( \frac{1}{c} \right).
\label{eq:c1c}
\end{equation}

\noindent
This result reflects the fact that in the dilute network limit
the first hitting time is dominated by the backtracking mechanism.

In the dense-network limit, where
$1.63 \sqrt{N} \ll c \le N$
the argument of the error function 
in Eq. (\ref{eq:T_FHT1})
satisfies
$\alpha \beta/ \sqrt{2} \ll 1$. 
In this case, one can use the approximation
${\rm erf}(x) \simeq x$,
which is the leading term in the Taylor expansion
of ${\rm erf}(x)$ around $x=0$.
Using this approximation, we obtain

\begin{equation}
\langle T_{\rm FH} \rangle \simeq  \sqrt{ \frac{\pi N}{2} }.
\end{equation}

\noindent
In this limit the first hitting time is dominated by the retracing mechanism.

\begin{figure}
\centerline{
\includegraphics[width=8cm]{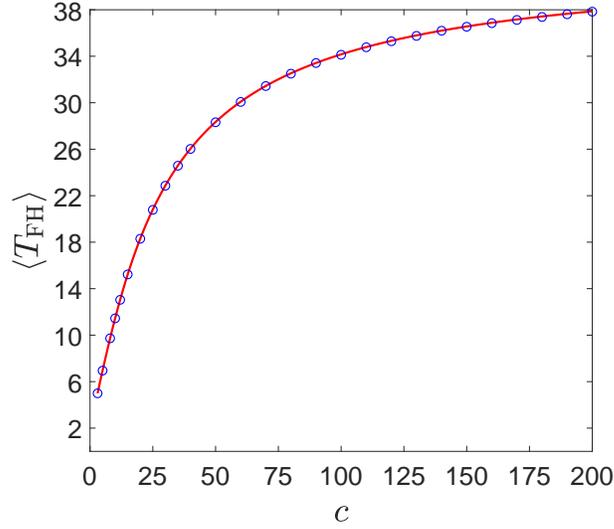}
}
\caption{
Analytical results for the
mean first hitting time 
$\langle T_{\rm FH} \rangle$ 
of RWs on RRGs
of size $N=1000$
(solid line),
as a function of the degree $c$.
The analytical results, 
obtained from Eqs.
(\ref{eq:<TFH>}),
(\ref{eq:T_FHT1}) 
and
(\ref{eq:T_FHT2})
are in excellent agreement with 
the results obtained from computer simulations
(circles).
}
\label{fig:5}
\end{figure}

In Fig. \ref{fig:5} 
we present 
analytical results for 
the mean first hitting time 
$ \langle T_{\rm FH} \rangle$ (solid line)
of RWs on RRGs
as a function of the degree $c$.
The analytical results, 
obtained from Eqs.
(\ref{eq:<TFH>}),
(\ref{eq:T_FHT1}) 
and
(\ref{eq:T_FHT2})
are in excellent agreement with 
the results obtained from computer simulations
(circles).
In the dilute-network limit $\langle T_{\rm FH} \rangle$ 
quickly increases as $c$ is increased, 
reaching saturation in the dense-network limit.

\section{The variance of the distribution of first hitting times}

Inserting $r=2$ in Eq. (\ref{eq:FH_r}) we obtain the
second moment
$\langle T_{\rm FH}^2 \rangle$ 
of the distribution of first hitting times,
which is given by

\begin{equation}
\langle T_{\rm FH}^2 \rangle = \sum_{t=0}^{N} 
(2 t + 1) P(T_{\rm FH}>t).
\label{eq:FH_sum2}
\end{equation} 

\noindent
Inserting the tail distribution $P(T_{\rm FH}>t)$ from Eq. (\ref{eq:FH_tail7})
into Eq. (\ref{eq:FH_sum2}),
shifting the summation index from $t$ to $t-2$
and replacing the sum by a sum of two integrals,
we obtain

\begin{equation}
\langle T_{\rm FH}^2 \rangle = J_1 + J_2,
\label{eq:<T2FH>}
\end{equation}

\noindent
where

\begin{eqnarray}
J_1 =
9 + 7 e^{- \beta}
+ \int_{3/2}^{N-3/2}
(2t+5) \exp \left[ - \frac{t(t-1)}{2 \alpha^2}
- 
\beta t \right] dt,
\label{eq:J1}
\end{eqnarray}

\noindent
and

\begin{eqnarray}
J_2 =
- \int_{3/2}^{N-3/2}
(2t+5)
\frac{ t(t-1)(2t-1)}{12 \left( \frac{c+2}{c-2} \right) \alpha^4}  
\exp \left[-\frac{  t   ( t-1 )}{2  \alpha^2}
- 
\beta
t \right] dt.
\label{eq:J2}
\end{eqnarray}

\noindent
Carrying out the integral in Eq. (\ref{eq:J1}) and taking the large network limit,
we obtain

\begin{eqnarray}
J_1
&\simeq&
9 + 7 e^{-\beta}
+
2 \alpha^2   \exp \left( -   \frac{20 \alpha^2 \beta + 15}{8 \alpha^2} \right)
\nonumber \\
&-& 
\alpha \sqrt{2 \pi} (\alpha^2 \beta - 3)
e^{  \frac{ (2 \alpha^2 \beta - 1)^2 }{8 \alpha^2}  }
\left[ 1 -
{\rm erf} \left( \frac{ \alpha^2 \beta + 1 }{ \sqrt{2} \alpha } \right)
\right].
\label{eq:TFH2c}
\end{eqnarray}

\noindent
Carrying out the integration in Eq. (\ref{eq:J2}) and taking the large network limit,
we obtain

\begin{eqnarray}
J_2 &\simeq&
 \frac{1}{12} 
\left( \frac{ c+2 }{c-2} \right)
 \left[4 \alpha ^4 \beta ^3-4 \alpha ^2 \beta  (4 \beta  -5)+15 \beta -36\right]
e^{-\frac{3}{8 \alpha ^2}-\frac{3 \beta }{2}}
\nonumber \\
&-&
\frac{1}{12} \sqrt{\frac{\pi }{2}}
\left( \frac{ c+2 }{c-2} \right)
 \alpha  
\left[4  \alpha ^4 \beta ^4-12 \alpha ^2 (\beta -2) \beta ^2-\beta ^2-36 \beta +12 \right] 
\times
\nonumber \\
& & \ \ \ \  e^{\frac{\left(1-2 \alpha ^2 \beta \right)^2}{8 \alpha ^2}} 
\left[1-\text{erf}\left(\frac{\alpha ^2 \beta  +1}{\sqrt{2} \alpha }\right)\right].
\label{eq:TFH2d}
\end{eqnarray}

In the limit of dilute networks, where
$3 \le c \ll 1.63 \sqrt{N}$, the product $\alpha \beta$ 
satisfies $\alpha \beta \gg 1$.
In this limit the second moment can be approximated by

\begin{eqnarray}
\langle T_{\rm FH}^2 \rangle 
&\simeq&
9 + 7 e^{-\beta}
- 
  \sqrt{2 \pi}  \alpha^3 \beta  
e^{  \left( \frac{ \alpha \beta }{\sqrt{2}} \right)^2  }
\left[ 1 -
{\rm erf} \left( \frac{ \alpha^2 \beta + 1 }{ \sqrt{2} \alpha } \right)
\right]
\nonumber \\
&+&
2 \alpha^2   e^{-5 \beta / 2}.
\label{eq:TFH2cp}
\end{eqnarray}

In the dense network limit, where
$1.63 \sqrt{N} \ll c < N$,
the parameter $\beta$ satisfies
$\beta \simeq 1/c$ and $\alpha \beta \ll 1$.
As a result, the error function on the right hand side of Eq. (\ref{eq:TFH2c})
becomes negligible 
and the second moment 
converges towards

\begin{equation}
\langle T_{\rm FH}^2 \rangle \simeq
2N. 
\end{equation}

The variance of the distribution of first hitting times is given by

\begin{equation}
{\rm Var}(T_{\rm FH}) =
\langle T_{\rm FH}^2 \rangle - \langle T_{\rm FH} \rangle^2.
\label{eq:Var}
\end{equation}

\noindent
In the dense network limit, the variance 
converges towards

\begin{equation}
{\rm Var}(T_{\rm FH}) \simeq 
\left( 2 - \frac{\pi}{2} \right) N.
\end{equation}

\begin{figure}
\centerline{
\includegraphics[width=8cm]{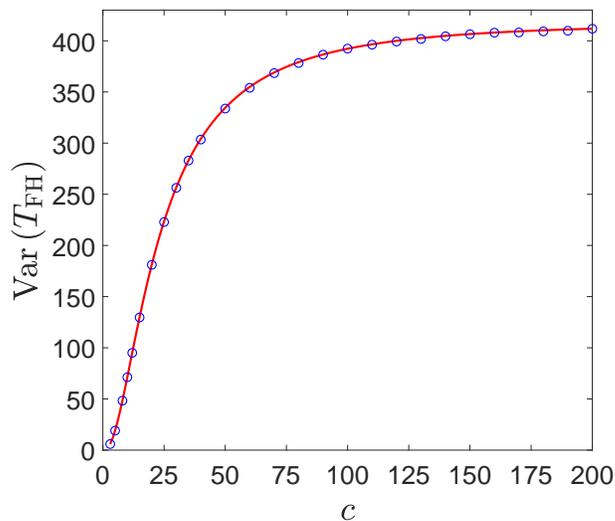}
}
\caption{
Analytical results for the
variance 
${\rm Var}(T_{\rm FH})$
of the distribution of first hitting times
of RWs on RRGs
of size $N=1000$
(solid line),
as a function of the degree $c$.
The analytical results, 
obtained from Eq.
(\ref{eq:Var}),
are in excellent agreement with 
the results obtained from computer simulations
(circles).
}
\label{fig:6}
\end{figure}

In Fig. \ref{fig:6} we present 
analytical results for the
variance of the distribution of first hitting times
of RWs on RRGs
of size $N=1000$
(solid line),
as a function of the degree $c$.
The analytical results, 
obtained from Eq.
(\ref{eq:Var}),
where $\langle T_{\rm FH}^2 \rangle$
is given by Eqs. 
(\ref{eq:<T2FH>}), 
(\ref{eq:TFH2c})
and
(\ref{eq:TFH2d})
and $\langle T_{\rm FH} \rangle$ 
is given by Eqs.
(\ref{eq:<TFH>}),
(\ref{eq:T_FHT1}) 
and
(\ref{eq:T_FHT2}),
are in excellent agreement with 
the results obtained from computer simulations
(circles).

\section{Analysis of the backtracking and retracing mechanisms}

The first hitting process consists of two competing scenarios,  
backtracking and retracing. In each instance of an RW trajectory
the first hitting time is determined by the scenario that occurs first.
We denote by $P_{\rm BT}$ the probability that the first hitting event of an RW starting from
a random initial node will take place by backtracking. 
Similarly, we denote by $P_{\rm RET}$
the probability that the first hitting event will take place by retracing. 
Since these are the only possible
mechanisms of first hitting of RWs, these two probabilities must satisfy
$P_{\rm BT} + P_{\rm RET}=1$.
The conditional probability that the first hitting 
event will take place at time $t$, given that it occurs via the backtracking scenario,
is denoted by
$P(T_{\rm FH}=t|{\rm BT})$.
Similarly, the conditional probability that the first hitting event will take place at 
time $t$ given that it occurs via the retracing scenario is denoted by
$P(T_{\rm FH}=t|{\rm RET})$.
The overall distribution of first hitting times 
can be expressed as a weighted sum of the two conditional distributions,
in the form

\begin{equation}
P(T_{\rm FH}=t) =
P_{\rm BT} P(T_{\rm FH}=t|{\rm BT})
+
P_{\rm RET}  P(T_{\rm FH}=t|{\rm RET}).
\label{eq:PTFHt}
\end{equation}
 
\noindent
The first term on the right hand side of Eq. (\ref{eq:PTFHt})
can be written in the form

\begin{equation}
P_{\rm BT} P(T_{\rm FH}=t|{\rm BT})
=
P(T_{\rm FH}>t-1) [ 1 - P_t(\lnot {\rm BT}) ],
\label{eq:PBTt}
\end{equation}

\noindent
namely as the probability that the first hitting event will not take place
up to time $t-1$ and will occur at time $t$ due to backtracking.
Inserting $P_t(\lnot {\rm BT})$ from Eq. (\ref{eq:PFHb}) into Eq. (\ref{eq:PBTt})
and summing up over $t$, it is found that the overall probability that the
first hitting event will take place via the backtracking scenario is given by

\begin{equation}
P_{\rm BT} = \frac{1}{c} \sum_{t=3}^{N+1} P(T_{\rm FH}>t-1).
\end{equation}

\noindent
Comparing with the tail-sum formula (\ref{eq:FH_sum}), 
the probability $P_{\rm BT}$ can be expressed in terms of the
mean first hitting time, namely

\begin{equation}
P_{\rm BT} = \frac{ \langle T_{\rm FH} \rangle - 2 }{c}.
\label{eq:PBT}
\end{equation}

\noindent
Thus, the complementary probability that the first hitting will occur via the retracing mechansim
is given by

\begin{equation}
P_{\rm RET} = 1 - \frac{ \langle T_{\rm FH} \rangle - 2 }{c}.
\label{eq:PRET}
\end{equation}

\noindent
Note that Eqs. (\ref{eq:PBT}) and (\ref{eq:PRET}) are exact.
However, inserting $\langle T_{\rm FH} \rangle$ from Eq. (\ref{eq:<TFH>})
with $I_1$ and $I_2$ given by Eqs. (\ref{eq:T_FHT1}) and (\ref{eq:T_FHT2}), 
respectively, the results become approximate due to the replacement of
the sum in Eq. (\ref{eq:FH_sum}) by integrals.

\begin{figure}
\centerline{
\includegraphics[width=8cm]{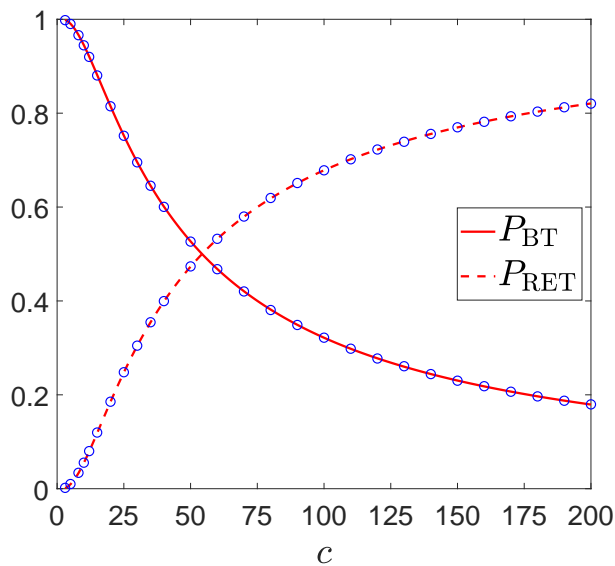}
}
\caption{
Analytical results for the
probabilities $P_{\rm BT}$ and $P_{\rm RET}$ that the first hitting event will
occur via the backtracking or the retracing scenarios, respectively.
The analytical results,
obtained from Eqs.
(\ref{eq:PBT}) and (\ref{eq:PRET}),
are in excellent agreement with 
the results obtained from computer simulations
(circles).
}
\label{fig:7}
\end{figure}

In Fig. \ref{fig:7} we present the probability $P_{\rm BT}$
that the first hitting event will occur via the backtracking scenario
and the complementary probability $P_{\rm RET}$ that
it will occur via the retracing scenario, as a function of the
degree $c$.
As expected, the probability $P_{\rm BT}$ decreases as
$c$ is increased while the probability $P_{\rm RET}$
increases. The crossover from the backtracking-dominated regime
of dilute networks
to the retracing-dominated regime of dense networks
occurs where $P_{\rm BT}=P_{\rm RET}=1/2$.
Using Eqs. (\ref{eq:PBT}) and (\ref{eq:PRET}) 
we find that at the crossover point
$\langle T_{\rm FH} \rangle = (c+4)/2$.
Comparing the right hand side of Eq. (\ref{eq:T_FHT1})
to $(c+4)/2$, it is found that the crossover point occurs at
$c \simeq 1.63 \sqrt{N}$.

\begin{figure}
\centerline{
\includegraphics[width=9cm]{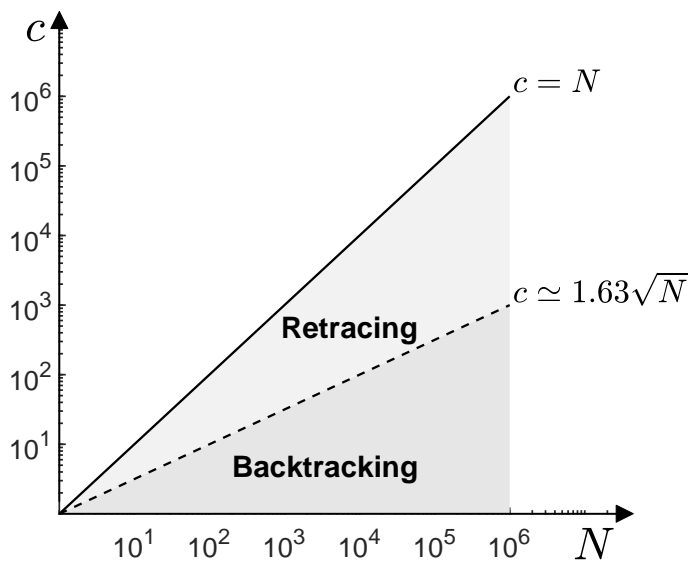}
}
\caption{
A diagram describing the first hitting process of RWs on RRGs.
In the regime of dilute networks, where $c \lesssim 1.63 \sqrt{N}$ the first hitting process
is dominated by backtracking, while in the regime of dense networks, where
$c \gtrsim 1.63 \sqrt{N}$ the first hitting process is dominated by retracing.
}
\label{fig:8}
\end{figure}

In Fig. \ref{fig:8} we present 
a diagram describing the first hitting process of RWs on RRGs.
In the regime of dilute networks, where $c \lesssim 1.63 \sqrt{N}$ the first hitting process
is dominated by backtracking, while in the regime of dense networks, where
$c \gtrsim 1.63 \sqrt{N}$ the first hitting process is dominated by retracing.

The conditional probability
$P(T_{\rm FH}=t|{\rm BT})$, for $t \ge 3$, 
can be written in the form

\begin{equation}
P(T_{\rm FH}=t|{\rm BT}) = 
\frac{ P(T_{\rm FH}>t-1) [ 1 - P_t(\lnot {\rm BT}) ] }{ P_{\rm BT} }.
\end{equation}

\noindent
Inserting $P_t(\lnot {\rm BT})$ from Eq. (\ref{eq:PFHb}) and
$P_{\rm BT}$ from Eq. (\ref{eq:PBT}), we obtain

\begin{equation}
P(T_{\rm FH}=t|{\rm BT}) = 
\frac{ P(T_{\rm FH}>t-1) }{ \langle T_{\rm FH} \rangle - 2 }.
\label{eq:PTFHtBT}
\end{equation}

\noindent
Focusing on the retracing mechanism,
the second term on the right hand side of Eq. (\ref{eq:PTFHt})
can be expressed in the form

\begin{equation}
P_{\rm RET} P(T_{\rm FH}=t|{\rm RET})
=
P(T_{\rm FH}>t-1) 
P_t(\lnot {\rm BT}) [ 1 - P_t(\lnot {\rm RET}| \lnot {\rm BT}) ].
\label{eq:PRETt}
\end{equation}

\noindent
Inserting 
$P_t(\lnot {\rm BT})$
from Eq. (\ref{eq:PFHb}),
$P_t(\lnot {\rm RET}| \lnot {\rm BT})$
from Eq. (\ref{eq:PFHr}), 
and $P_{\rm RET}$ from Eq. (\ref{eq:PRET})
into Eq. (\ref{eq:PRETt}),
we obtain

\begin{equation}
P(T_{\rm FH}=t|{\rm RET})
=
\left( \frac{  c - 1  }{ c -   \langle T_{\rm FH} \rangle + 2 } \right)
  \frac{(c-2)(t-3)+1}{(N-2)c-2(t-3)+1}  
P(T_{\rm FH}>t-1).
\label{eq:PTfhtret}
\end{equation}

\begin{figure}
\centerline{
\includegraphics[width=14cm]{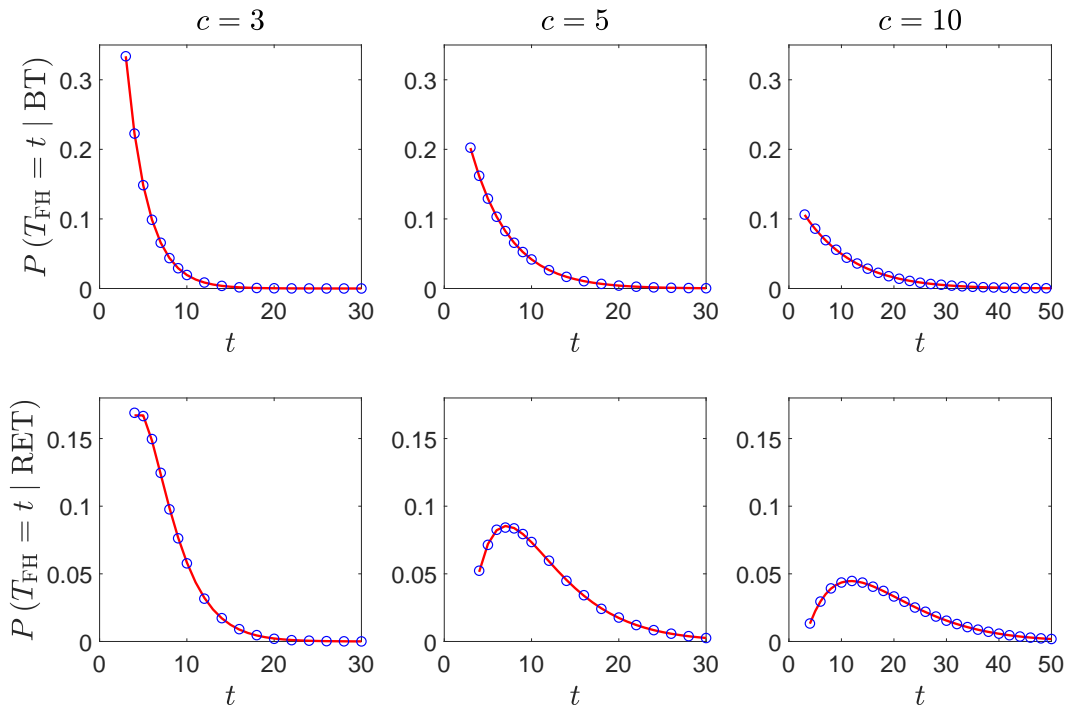}
}
\caption{
Analytical results for the
distribution $P(T_{\rm FH}=t | {\rm BT})$ and 
$P(T_{\rm FH} = t | {\rm RET})$ of first hitting  
times, conditioned on the first hitting to 
occur via the backtracking or the retracing scenarios,
respectively.
The analytical results, obtained from Eqs.
(\ref{eq:PTFHtBT}) and (\ref{eq:PTfhtret}),
are in excellent agreement with 
the results obtained from computer simulations
(circles).
}
\label{fig:9}
\end{figure}

In Fig. \ref{fig:9} we present the probabilities 
$P(T_{\rm FH}=t|{\rm BT})$
and 
$P(T_{\rm FH}=t|{\rm RET})$
that the first hitting process will take place at time $t$,
under the condition that it occurs via the backtracking and the retracing scenarios, respectively.

The mean first hitting time under the condition that the first hitting occurs via the
backtracking scenario is given by

\begin{equation}
\mathbb{E}[T_{\rm FH} | {\rm BT}] = 
\sum_{t=3}^{N+1} t P(T_{\rm FH}=t|{\rm BT}).
\label{eq:EFHBT}
\end{equation}

\noindent
Inserting 
$P(T_{\rm FH}=t|{\rm BT})$
from Eq. (\ref{eq:PTFHtBT}) into Eq. (\ref{eq:EFHBT}), we obtain

\begin{equation}
\mathbb{E}[T_{\rm FH} | {\rm BT}] = 
\frac{1}{ \langle T_{\rm FH} \rangle - 2 }
\sum_{t=3}^{N+1}  
t P(T_{\rm FH}>t-1) 
\label{eq:EFHBT2}
\end{equation}

\noindent
Using the tail-sum formula
\cite{Pitman1993},
it is found that

\begin{equation}
\mathbb{E}[T_{\rm FH} | {\rm BT}] = 
 \frac{\langle T_{\rm FH}^2 \rangle + \langle T_{\rm FH} \rangle - 6}{ 2( \langle T_{\rm FH} \rangle - 2 ) }.
\label{eq:EFHBT2}
\end{equation}

The mean first hitting time $\langle T_{\rm FH} \rangle$ can be expressed as
a weighted sum of the form

\begin{equation}
\langle T_{\rm FH} \rangle = 
\mathbb{E}[T_{\rm FH}|{\rm BT}] P_{\rm BT}
+
\mathbb{E}[T_{\rm FH}|{\rm RET}] P_{\rm RET}.
\label{eq:Tfhws}
\end{equation}

\noindent
Inserting $\mathbb{E}[T_{\rm FH}|{\rm BT}]$ from Eq. (\ref{eq:EFHBT2}) 
into Eq. (\ref{eq:Tfhws}) 
and solving for
$\mathbb{E}[T_{\rm FH}|{\rm RET}]$,
we obtain

\begin{equation}
\mathbb{E}[T_{\rm FH}|{\rm RET}]=
\frac{  \langle T_{\rm FH}^2 \rangle - (2c-1) \langle T_{\rm FH} \rangle - 6 }
{ 2 ( \langle T_{\rm FH} \rangle  - c - 2 ) }.
\label{eq:EFHRET2}
\end{equation}

\begin{figure}
\centerline{
\includegraphics[width=8cm]{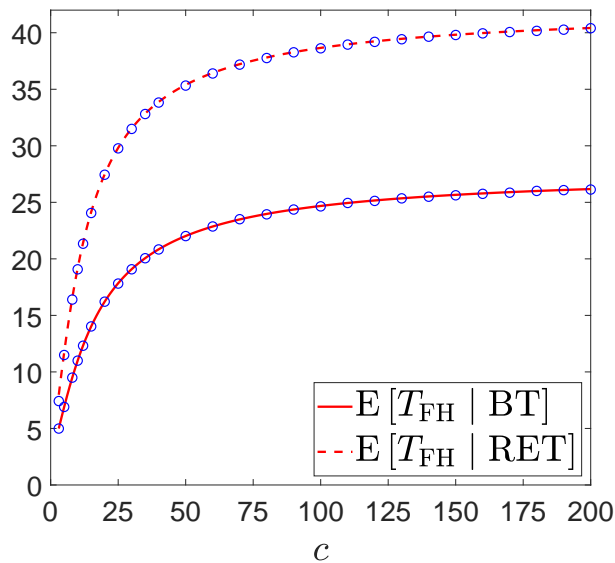}
}
\caption{
Analytical results for the
conditional expectation values
$\mathbb{E}[T_{\rm FH} | {\rm BT}]$
and
$\mathbb{E}[T_{\rm FH} | {\rm RET}]$
of the first hitting time, vs. the degree $c$
given that the first hitting event occurs via the
backtracking or the retracing scenario, respectively.
The analytical results, obtained from Eqs.
(\ref{eq:EFHBT2}) and (\ref{eq:EFHRET2}),
are in excellent agreement with 
the results obtained from computer simulations
(circles).
}
\label{fig:10}
\end{figure}

In Fig. \ref{fig:10} we present 
analytical results for the
conditional expectation values
$\mathbb{E}[T_{\rm FH} | {\rm BT}]$
and
$\mathbb{E}[T_{\rm FH} | {\rm RET}]$
of the first hitting time, vs. the degree $c$
given that the first hitting event occurs via the
backtracking or the retracing scenario, respectively.
The analytical results, obtained from Eqs.
(\ref{eq:EFHBT2}) and (\ref{eq:EFHRET2}),
are in excellent agreement with 
the results obtained from computer simulations
(circles).

\section{Comparing $P(T_{\rm FH}>t)$ between RWs on RRGs and RWs on ER networks}

The distribution $P(T_{\rm FH}>t)$ of first hitting times of RWs on RRGs
is given by Eq. (\ref{eq:FH_tail}).
It can be expressed as a product of 
a geometric distribution, 
associated with the backtracking process,
and a Rayleigh distribution, associated with the retracing process.
The Rayleigh distribution is parameterized by the parameter $\alpha$,
while the geometric distribution is parameterized by $\beta$.
It is interesting to compare the results obtained above for the distribution
$P(T_{\rm FH}>t)$ 
of RWs on RRGs
of size $N$ and degree $c$  
with the corresponding results 
for RWs on ER networks of the same size and mean degree $\langle K \rangle=c$  
\cite{Tishby2017}.
Another interesting comparison is between the distributions of first hitting times
of NBWs on RRGs
and ER networks.
In all these cases, the tail distribution of first hitting times is described by
the same functional form as in
Eq. (\ref{eq:FH_tail}), but the expressions for $\alpha$ and $\beta$ 
are different.
In Table 1 we present a $2 \times 2$ diagram that provides the parameters
$\alpha$ and $\beta$ for RWs on RRGs
(upper-left cell),
NBWs on RRGs
(upper-right cell),
RWs on ER networks (lower-left cell) and NBWs on ER networks (lower-right cell).

\begin{table}
\caption{The parameters $\alpha$ and $\beta$ of the distribution $P(T_{\rm FH}>t)$ 
[Eq. (\ref{eq:FH_tail})]
for RWs and NBWs on RRGs and on ER networks}
\begin{center}
\begin{tabular}{| l |  l  | l  | }
 \hline \hline
              &           &        \\
          & \ \ \  RW  & \ \ \  NBW    \\ 
            &           &        \\
     \hline \hline 
        &        & \\
      &  $\alpha = \sqrt{ \left( \frac{c}{c-2} \right) N }$   &   $\alpha = \sqrt{ \left(  \frac{c }{c-2} \right)  N}$    \\  
 \ \     \rotatebox{90}{RRG}    \ \       &                                                            &                                              \\
             &            $\beta   = \ln \left( \frac{c}{c-1} \right)$                 &    $\beta=0$                    \\ 
         &        & \\             \hline
        &        & \\
       &   $\alpha = \sqrt{\left( \frac{c+1 }{c} \right) N}$   &    $\alpha = \sqrt{ \left( \frac{c+1 }{c} \right) N }$     \\  
 \ \ \ \  \rotatebox{90}{ER}     \ \      &             &                                             \\
           &      $\beta = \ln \left( \frac{c}{c-1+e^{-c}} \right)$                     &     $\beta=- \ln (1-e^{-c})$           \\
        &        & \\ \hline
\hline
\hline \hline
\end{tabular}
\end{center}
\label{table}
\end{table}

One difference between RRGs
and ER networks is that ER networks include isolated nodes of degree $k=0$.
Since an RW starting from an isolated node cannot make even a single move,
the initial node of an RW on an ER network is chosen randomly from all the nodes of degree $k \ge 1$.
Moreover, isolated nodes are not accessible to the RW at times $t>1$.
As a result, for RWs on ER networks the probability of backtracking at any time $t$
is given by $P_t({\rm BT})=(1-e^{-c})/c$, compared to $1/c$ in the case of RWs
on RRGs
\cite{Tishby2017}.
This is reflected in the parameter $\beta$ for RWs on ER networks, which includes
the $e^{-c}$ term.

The effect of the retracing process on the distribution of first hitting times
is described by a Rayleigh distribution parametrized by $\alpha$.
In fact, the probability that at any time $t$ the RW will hop into 
any one of the nodes that were previously visited up to time $t-3$
is given by $1/\alpha^2$. 
Naively, one expects that the probability to enter a specific node at time $t$ is $1/N$.
However, each visit of a node exhausts two of its edges, which cannot be used
to revisit the node in the retracing scenario.
Therefore, the probability that at a given time an RW will revisit a specific node
that has already been visited two or more time steps earlier
is smaller than $1/N$. In the case of RWs on RRGs
it is given by
$(1-2/c)/N$, while in the case of RWs on ER networks it is given by
$[1-1/(c+1)]/N$. 
These corrections are reflected in the values of $\alpha$ in Table 1.

\begin{figure}
\centerline{
\includegraphics[width=8cm]{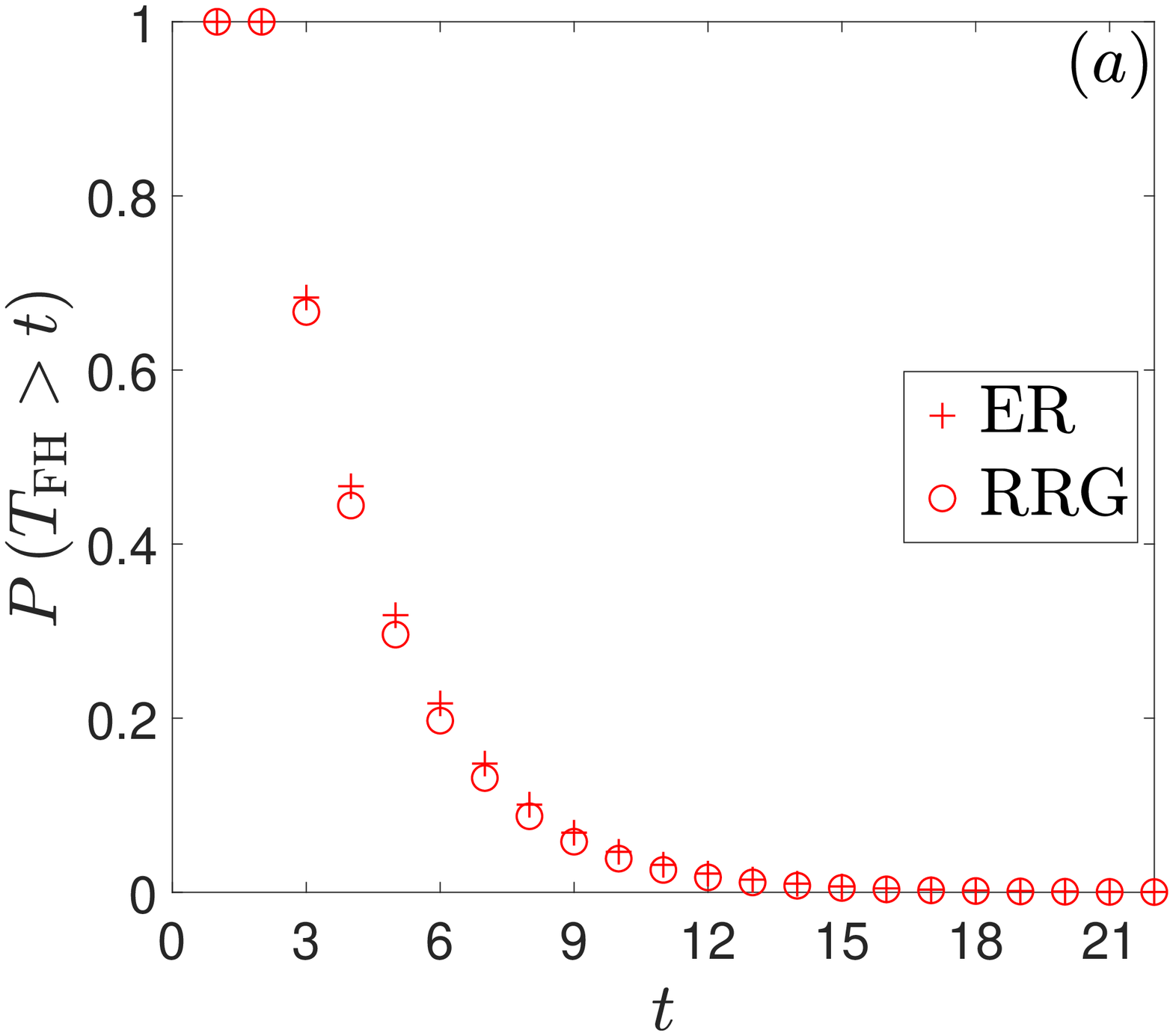}
}
\centerline{
\includegraphics[width=8cm]{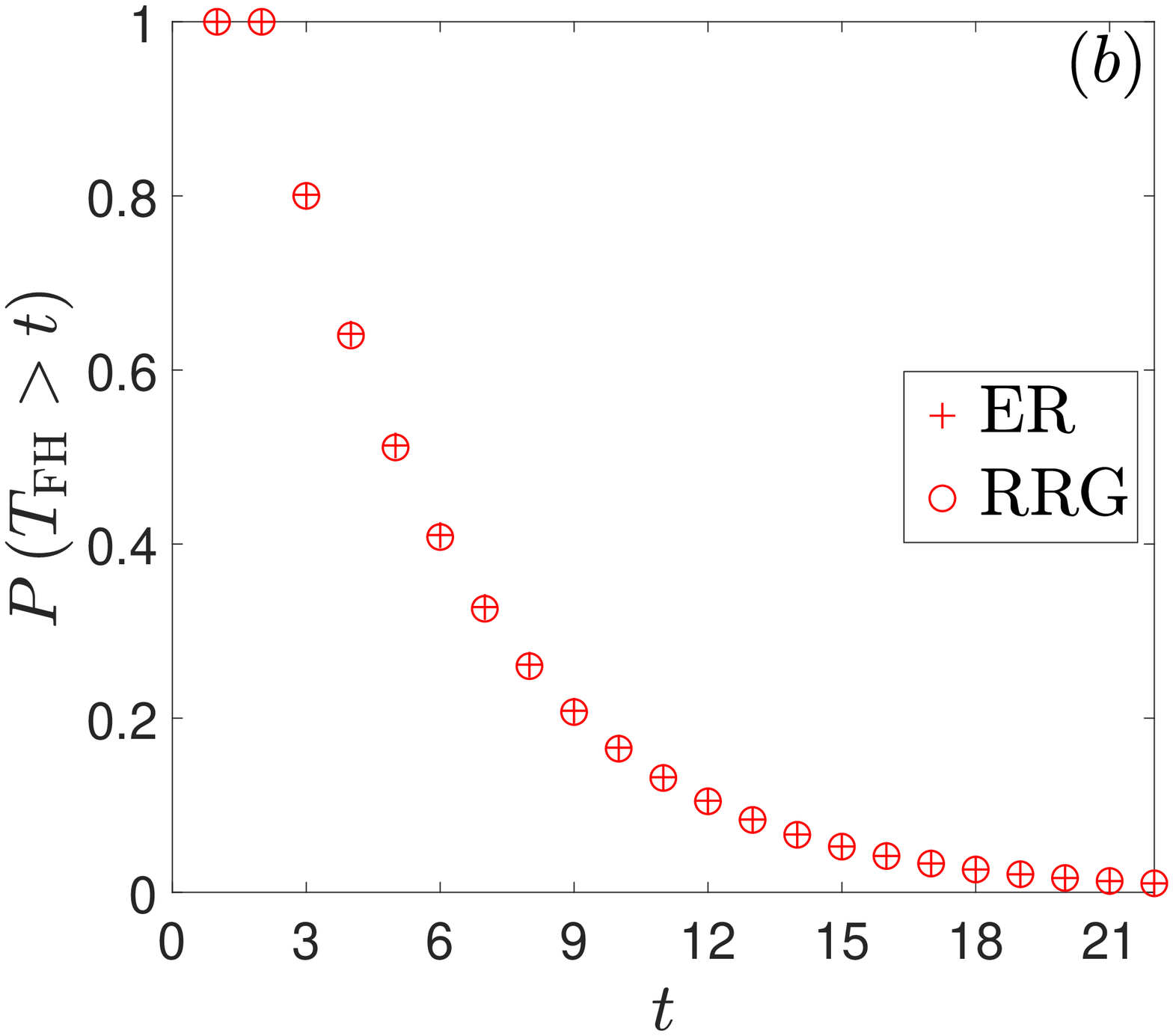}
}
\caption{
Comparison between the
analytical results for the 
tail distributions 
$P(T_{\rm FH} > t)$  
of first hitting times  
of RWs on RRGs ($\circ$) 
and RWs on ER networks  ($+$).
The network size is $N=1000$.
The degree is
(a) $c=3$ and (b) $c=5$.
In both cases, the tail distribution takes the form of Eq. (\ref{eq:FH_tail}),
with the parameters $\alpha$ and $\beta$ given in Table 1.
Note that the agreement between the analytical results and computer simulations
was established in Fig. \ref{fig:3} (for RWs on RRGs) and in Ref. \cite{Tishby2017}
(for RWs on ER networks).
}
\label{fig:11}
\end{figure}

In Fig. \ref{fig:11} we compare the analytical results for the distributions $P(T_{\rm FH}>t)$
of RWs on RRGs
(circles) and ER networks ($+$) 
of size $N=1000$ and (a) $c=3$ and (b) $c=5$.
For $c=3$, where the first hitting process is dominated by the
backtracking scenario, there is a slight difference at short times due to $e^{-c}$
term in the expression for $\beta$ in the ER network.
For $c=5$ the results obtained for the two networks are found to be similar.
This is due to the fact that as $c$ is increased the
$e^{-c}$ term in the expression for $\beta$ of the ER network
becomes negligible and it converges towards the
expression for $\beta$ of the RRG.
This convergence occurs within the dilute network regime, in which the
first hitting process is dominated by backtracking.

In order to take a closer look at the retracing process, we consider the 
distribution of first hitting times in NBWs, in which the backtracking process
is suppressed. In the case of NBWs on RRGs,
upon suppression
of the backtracking process, the retracing scenario remains unchanged and
the distribution of first hitting times becomes a Rayleigh distribution.
In contrast, in the case of NBWs on ER networks the suppression of the
backtracking process gives rise to a new mechanism of first hitting,
referred to as the trapping scenario. This scenario occurs when the
NBW enters a leaf node of degree $k=1$. In the following time step it 
becomes trapped in the leaf node because the backtracking move
into the previous node is not allowed. The distribution
of first hitting times of NBWs on ER networks was studied in Ref. \cite{Tishby2017b}
and the corresponding expressions for $\alpha$ and $\beta$
are shown in the lower-right cell in Table 1.

\begin{figure}
\centerline{
\includegraphics[width=8cm]{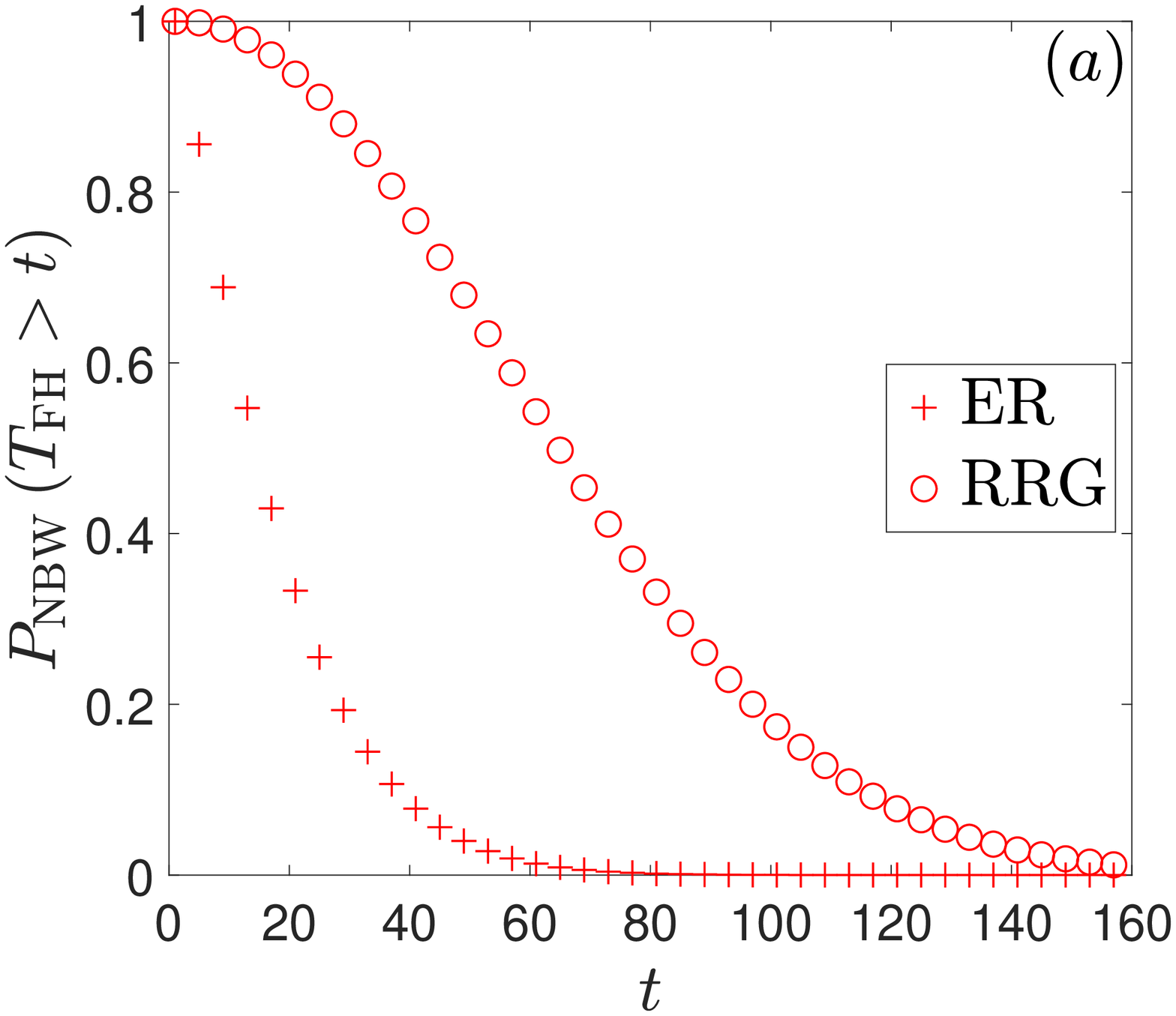}
}
\centerline{
\includegraphics[width=8cm]{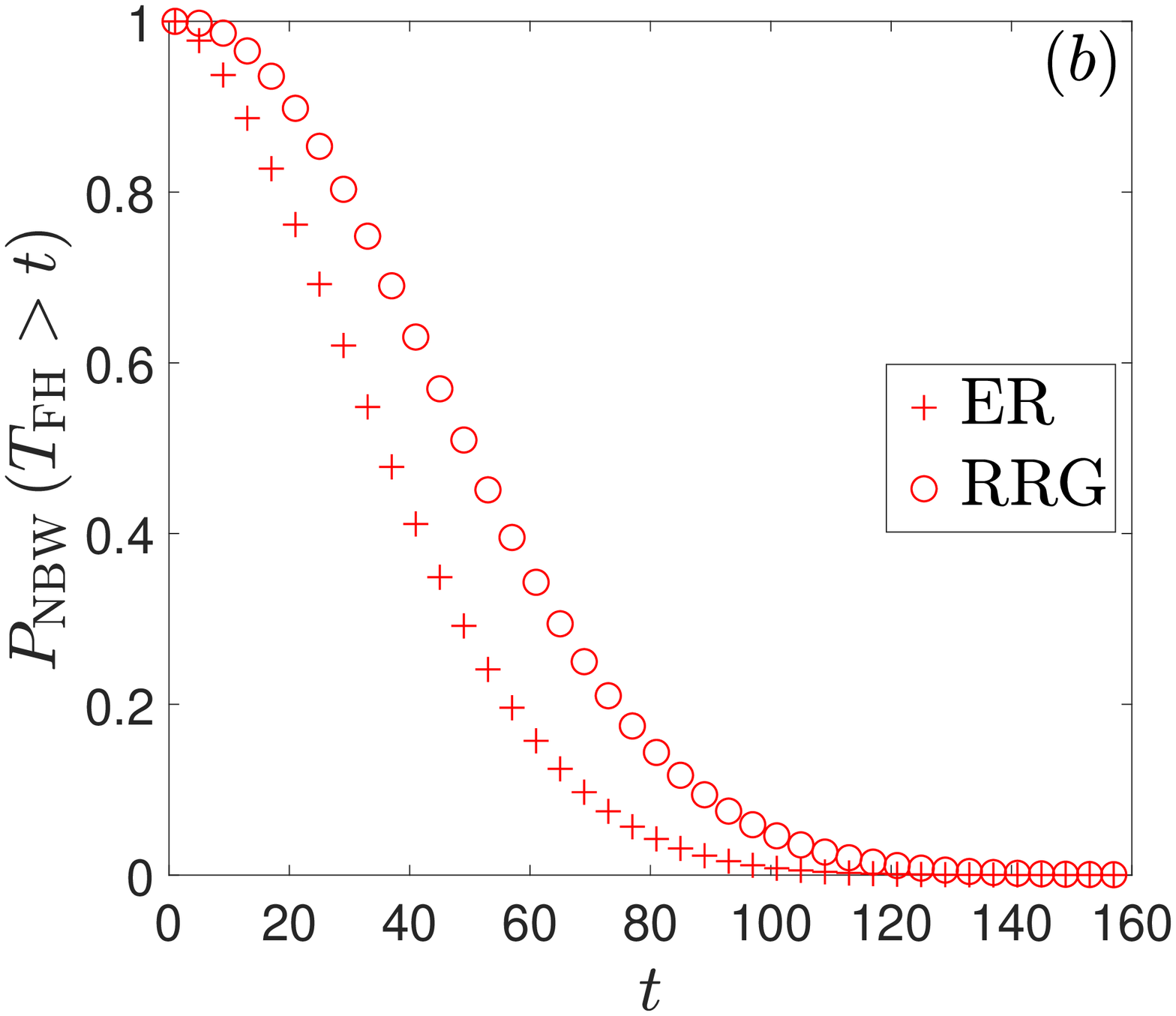}
}
\caption{
Comparison between the
analytical results for the tail distributions 
$P_{\rm NBW}( T_{\rm FH} > t )$  
of first hitting times  
of NBWs on RRGs
($\circ$) and NBWs on ER networks ($+$).
The network size is of size $N=1000$.
The degree is
(a) $c=3$ and (b) $c=5$.
In both cases, the tail distribution takes the form of Eq. (\ref{eq:FH_tail}),
with the parameters $\alpha$ and $\beta$ given in Table 1.
Note that the agreement between the analytical results and computer simulations
was established in Fig. \ref{fig:4} (for NBWs on RRGs) and in Ref. \cite{Tishby2017b}
(for NBWs on ER networks).
}
\label{fig:12}
\end{figure}

In Fig. \ref{fig:12} we present
a comparison between
analytical results for the distributions $P_{\rm NBW}( T_{\rm FH}>t )$
of NBWs on RRGs
(circles) and ER networks ($+$) 
of size $N=1000$ and (a) $c=3$ and (b) $c=5$.
The first hitting times of NBWs on ER networks are found to be much shorter than
those obtained for NBWs on RRGs.
This contrast is most pronounced for small values of $c$.
This is due to the emergence of the trapping scenario which is
most effective in the limit of dilute networks.

\section{Discussion}

Beyond the specific problem of first hitting times of RW on networks,
the analysis presented here provides useful insight into the general
context of the distribution of life expectancies of humans, animals and machines
\cite{Finkelstein2008,Gavrilov2001}.
It illustrates the combination of two lethal hazards, where one hits at a fixed,
age-independent rate, while the other increases linearly with age.
The first hazard may be considered as an external cause such as an accident
while the second hazard involves some aging related 
degradation which results
in an increasing failure rate.

In a more specific context of survival problems, the 
RW model that terminates upon its first hitting event
can be cast in 
the language of foraging theory
as a model describing a wild animal, which is 
randomly foraging in a random
network environment
\cite{Chupeau2016}. 
Each time the animal visits 
a node it consumes all the food
available in this node and needs to move on to 
one of the adjacent nodes.
The model describes rather harsh conditions, in 
which the regeneration of 
resources is very slow and the visited nodes do 
not replenish within the lifetime
of the forager. Moreover, the forager does not 
carry any reserves and in order
to survive it must hit a vital node at every time step. 
More realistic variants of this model have been 
studied on lattices of 
different dimensions. It was shown 
that under slow regeneration
rates, the forager is still susceptible to starvation, 
while above some threshold of the regeneration 
rate, the probability of starvation diminishes significantly
\cite{Chupeau2016}. 
The case in which the forager
carries sufficient resources that enable it to avoid starvation even when it 
visits up to $S$ non-replenished nodes in a row, was also studied 
\cite{Benichou2014,Benichou2016}.

In the dense network limit
the first hitting process is dominated by the retracing scenario.
As a result,   
the distribution of first hitting times becomes insensitive to the degree $c$.
In this limit the mean first hitting time
scales like
$\sqrt{N}$.
This can be understood as follows.
In this limit, the backtracking probability 
is very low and thus the backtracking-induced
first hitting events become negligible.
Instead, retracing becomes the dominant scenario. 
Due to the very high connectivity, the hopping between adjacent nodes
can be considered as the simple combinatorial problem of randomly choosing 
one node at a time from a set of $N$ nodes, 
allowing each node to be chosen more than once.
In this limit the statistical properties of first hitting times become
analogous to those of the birthday problem
\cite{Knight1973,Flajolet1992}.
More specifically, in this limit the probability that $P(T_{\rm FH}>t)$ in 
a network that consists of $N=365$ nodes is equal to the probability
that in a party of $t$ participants there will not be even one pair who
share the same birthday
\cite{Knight1973}.

\section{Summary}

We presented a statistical analysis of the first hitting times of
RWs on RRGs,
which may take place either via backtracking or via retracing.
The tail distribution 
$P( T_{\rm FH} > t )$
of first hitting times  was calculated.
It can be expressed as a product of a geometric distribution 
associated with the backtracking process
and a Rayleigh distribution which is due to the retracing process.
We also obtained closed form expressions for 
the mean first hitting time $\langle T_{\rm FH} \rangle$ and for the variance 
${\rm Var}(T_{\rm FH})$ of the distribution of first hitting times.
The analytical results are 
found to be in excellent agreement with the results 
obtained from computer simulations.
We obtained analytical results for the probabilities 
$P_{\rm BT}$ and $P_{\rm RET}$ that
the first hitting event will occur via the backtracking or retracing scenarios, respectively. 
We showed that in dilute networks the dominant first hitting scenario is backtracking while in
dense networks the dominant scenario is retracing. 
We also obtained expressions for
the conditional distributions of first hitting time, 
$P(T_{\rm FH}=t| {\rm BT})$ and 
$P(T_{\rm FH}=t| {\rm RET})$,
in which the first hitting event occurs via the backtracking or the retracing scenario, respectively. 
These results provide useful
insight into the general problem of survival analysis and the statistics of mortality
rates when two or more termination scenarios coexist.
We also analyzed the distribution of first hitting times in non-backtracking random walks (NBWs),
in which the backtracking process is suppressed and
compared the results obtained here for RWs and NBWs on RRGs to earlier results for RWs
and NBWs on Erd{\H o}s-R\'enyi networks.

This work was supported by the Israel Science Foundation grant no. 
1682/18.




\section*{References}

\begin{thebibliography}{10}

\bibitem{Spitzer1964}
Spitzer F 1964
{\it Principles of Random Walk}
(New York: Springer-Verlag)

\bibitem{Weiss1994}
Weiss G H 1994
{\it Aspects and Applications of the Random Walk}
(New York: North Holland)

\bibitem{Berg1993}
Berg H C 1993
{\it Random Walks in Biology}
(Princeton: Princeton University Press)

\bibitem{Ibe2013}
Ibe O C 2013
{\it Elements of Random Walk and Diffusion Processes}
(New Jersey: Wiley \& Sons)

\bibitem{Fisher1966}
Fisher M E 1966
Shape of a self‐avoiding walk or polymer chain
{\it J. Chem. Phys.} {\bf 44} 616

\bibitem{Edwards1965}
Edwards S F 1965
The statistical mechanics of polymers with excluded volume,
{\it Proceedings of the Physical Society} {\bf 85} 613


\bibitem{Degennes1979}
De Gennes P G 1979
{\it Scaling Concepts in Polymer Physics}
(Ithaca: Cornell University Press) 

\bibitem{Evans2011}
Evans M R and Majumdar S N 2011
Diffusion with stochastic resetting
{\it Phys. Rev. Lett.} {\bf 106} 160601 

\bibitem{Lopez2012} 
Lopez Mill\'an V M, Cholvi V, Lopez L and Anta A F 2012
A model of self‐avoiding random walks for searching complex networks
{\it Networks} {\bf 60} 71



\bibitem{Lawler2010b}
Lawler G F 2010
{\it Random Walk and the Heat Equation}
(Providence: American Mathematical Society)

\bibitem{Lawler2010a}
Lawler G F and Limic V 2010
{\it Random Walk: A Modern Introduction}
(Cambridge: Cambridge University Press)

\bibitem{ben-Avraham2000}
ben-Avraham D and Havlin S 2000 
{\it Diffusion and Reactions in Fractals and Disordered Systems} 
(Cambridge: Cambridge University Press)

\bibitem{Noh2004}
Noh D J and Rieger H 2004
Random walks on complex networks
{\it Phys. Rev. Lett.} {\bf 92} 118701 

\bibitem{Havlin2010}
Havlin S and Cohen R 2010
{\it Complex Networks: Structure, Robustness and Function}
(Cambridge University Press, New York) 

\bibitem{Newman2010}
Newman M E J 2018 {\it Networks: an Introduction},
2$^{\rm nd}$ Ed.
(Oxford: Oxford University Press) 



\bibitem{Pastor-Satorras2001}
Pastor-Satorras R and  Vespignani A 2001
Epidemic spreading in scale-free networks
{\it Phys. Rev. Lett.} {\bf 86} 3200 

\bibitem{Barrat2012}
Barrat A, Barth\'elemy M and Vespignani A 2012
{\it Dynamical Processes on Complex Networks}
(Boston: Cambridge University Press)


\bibitem{Debacco2015}
De Bacco C, Majumdar S N and Sollich P 2015 
The average number of distinct sites visited by a random walker on random graphs
{\it J. Phys. A} {\bf 48} 205004

\bibitem{Montroll1965}
Montroll E W and Weiss G H 1965
Random Walks on Lattices II
{\it J. Math. Phys.} {\bf 6} 167


\bibitem{Herrero2003}
Herrero C P and Saboy\'a M 2003
Self-avoiding walks and connective constants in small-world networks
{\it Phys. Rev. E} {\bf 68} 026106

\bibitem{Herrero2005b}
Herrero C P 2005
Kinetic growth walks on complex networks
{\it J. Phys. A} {\bf 38} 4349



\bibitem{Redner2001}
Redner S 2001 
{\it A Guide to First Passage Processes}
(Cambridge: Cambridge University Press)

\bibitem{HittingPassage}
Note that occasionally in the literature the term
"first hitting time" is used as an alternative to
the term "first passage time".




\bibitem{Kahn1989}
Kahn J D, Linial N, Nisan N and Saks M E 1989 
On the cover time of random walks on graphs 
{\it J. Theor. Probab.} {\bf 2} 121  



\bibitem{Alon2007}
Alon N, Benjamini I, Lubetzky E and Sodin S 2007
Non-backtracking random walks mix faster
{\it Commun. Contemp. Math.} {\bf 9} 585



\bibitem{Tishby2017}
Tishby I, Biham O and Katzav E 2017
The distribution of first hitting times of
random walks on Erd{\H o}s-R\'enyi networks
{\it J. Phys. A}  {\bf 50} 115001

\bibitem{Tishby2017b}
Tishby I, Biham O and Katzav E 2017
The distribution of first hitting times of nonbacktracking random walks on Erd{\H o}s-R\'enyi
networks
{\it J. Phys. A}  {\bf 50} 205003


\bibitem{Molloy1995}
Molloy M and Reed A 1995 
A critical point for random graphs with a given degree sequence
{\it Rand. Struct. Alg.} {\bf 6}  161 


\bibitem{Molloy1998}
Molloy M and Reed A 1998 
The Size of the Giant Component of a Random Graph with a Given Degree Sequence
{\it Combinatorics, Probability and Computing} {\bf 7}  295 


\bibitem{Newman2001}
Newman M E J, Strogatz S H and Watts D J 2001
Random graphs with arbitrary degree distributions and their applications,
{\it Phys. Rev. E} {\bf 64}  026118  



\bibitem{Erdos1959}
Erd{\H o}s P and R\'enyi A 1959 
On random graphs.  I
{\it Publicationes Mathematicae (Debrecen)}
{\bf 6}  290 

\bibitem{Erdos1960}
Erd{\H o}s P and R\'enyi A 1960
On the evolution of random graphs
{\it Publ. Math. Inst. Hung. Acad. Sci.} 
{\bf 5} 17

\bibitem{Erdos1961}
Erd{\H o}s P and R\'enyi A 1961
On the evolution of random graphs.  II
{\it Bull. Inst. Int. Stat.} 
{\bf 38} 343




 





\bibitem{Bonneau2017}
Bonneau H, Hassid A, Biham O, K\"uhn R and Katzav E 2017 
Distribution of shortest cycle lengths in random networks 
{\it Phys. Rev. E} {\bf 96} 062307 



\bibitem{Pathria2011}
Pathria R K and Beale P D 2011
{\it Statistical Mechanics 3rd Edition}
(Amsterdam: Academic Press) 


\bibitem{Papoulis2002}
Papoulis A, Pillai S and Unnikrishna S 2002 
{\it Probability, Random Variables and Stochastic Processes}
(Boston: McGraw-Hill) 


\bibitem{Pitman1993}
Pitman J 1993
{\it Probability} 
(New York: Springer-Verlag)


\bibitem{Olver2010}
Olver F W J, Lozier D M, Boisvert R R and Clark C W 2010
{\it NIST Handbook of Mathematical Functions}
(Cambridge: Cambridge University Press) 



\bibitem{Finkelstein2008}
Finkelstein M 2008
{\it Failure Rate Modeling for Reliability and Risk}
(London: Springer-Verlag) 

\bibitem{Gavrilov2001}
Gavrilov L A and Gavrilova N S 2001
The reliability theory of aging and longevity
{\it J. theor. Biol} {\bf 213} 527 

\bibitem{Chupeau2016}
Chupeau M, B\'enichou O and Redner S 2016
Universality classes of foraging with resource renewal
{\it Phys. Rev. E} {\bf 93} 032403

\bibitem{Benichou2014}
B\'enichou O and Redner S 2014
Depletion-Controlled Starvation of a Diffusing Forager
{\it Phys. Rev. Lett} {\bf 113} 238101

\bibitem{Benichou2016}
B\'enichou O, Chupeau M and Redner S 2016
Role of depletion on the dynamics of a diffusing forager
{\it J. Phys. A} {\bf 49} 394003



\bibitem{Knight1973}
Knight W and Bloom D M 1973 
A birthday problem
{\it Amer. Math. Monthly} {\bf 80} 1141



\bibitem{Flajolet1992}
Flajolet P, Gardy D and Thimonier L 1992
Birthday paradox, coupon collectors, caching algorithms and self-organizing search
{\it Discrete Applied Mathematics} {\bf 39} 207  


\end{thebibliography}
\end{document}